\documentclass{aa}  
\usepackage{tikz}
\usepackage{txfonts}
\usepackage{subcaption}

\usepackage{natbib,twoopt}
\usepackage[colorlinks,breaklinks=true]{hyperref}
\hypersetup{
    citecolor={blue},
    linkcolor={blue},
    urlcolor={blue}
}
\bibpunct{(}{)}{;}{a}{}{,}             
\makeatletter
  \newcommandtwoopt{\citeads}[3][][]{\href{http://adsabs.harvard.edu/abs/#3}%
    {\def\hyper@linkstart##1##2{}%
     \let\hyper@linkend\@empty\citealp[#1][#2]{#3}}}
  \newcommandtwoopt{\citepads}[3][][]{\href{http://adsabs.harvard.edu/abs/#3}%
    {\def\hyper@linkstart##1##2{}%
     \let\hyper@linkend\@empty\citep[#1][#2]{#3}}}
  \newcommandtwoopt{\citetads}[3][][]{\href{http://adsabs.harvard.edu/abs/#3}%
    {\def\hyper@linkstart##1##2{}%
     \let\hyper@linkend\@empty\citet[#1][#2]{#3}}}
  \newcommandtwoopt{\citeyearads}[3][][]%
    {\href{http://adsabs.harvard.edu/abs/#3}
    {\def\hyper@linkstart##1##2{}%
     \let\hyper@linkend\@empty\citeyear[#1][#2]{#3}}}
\makeatother

\newcommand{\kms}{\mbox{$\>{\rm km\, s^{-1}}$}}
\newcommand{\pc}{\>{\rm pc}}
\newcommand{\kpc}{\mbox{$\>{\rm kpc}$}} 

\newcommand{\Gyr}{\mbox{$\>{\rm Gyr}$}}
\newcommand{\Myr}{\mbox{$\>{\rm Myr}$}}

\newcommand{\Msun}{\>{\rm M_{\odot}}}

\newcommand\degree{^\circ}

\newcommand{\feh}{\mbox{$\rm [Fe/H]$}}

\newcommand\lv{{$l_{\rm v}$}}

\mathchardef\mhyphen="2D

\usepackage{soul}
\usepackage[normalem]{ulem}
\usepackage{xcolor}

\begin{document} 

   \title{Kinematic Diagnostics for Non-Axisymmetry in the Milky Way's Nuclear Stellar Disc}

   \author{K. Fiteni
          \inst{1,2}
          \and
          X. Li\inst{1}
          \and
          M. C. Sormani\inst{1}
          \and
           V. P. Debattista\inst{3}
          \and
          A. Vasini\inst{1}
          \and
          F. Nogueras-Lara\inst{4}
          \and
          J. L. Sanders\inst{5}
          \and
          N. Deg\inst{6}
          \and
          M. Schultheis\inst{7}
          \and
          M. Donati\inst{1}
          \and
          Z.-X. Feng\inst{1}
          }

   \institute{Como Lake centre for AstroPhysics (CLAP), DiSAT, Università dell’Insubria, Via Valleggio 11, 22100 Como, Italy\\
   \email{karlfiteni@gmail.com}
   \and
     Institute of Space Sciences \& Astronomy, University of Malta, Msida MSD 2080, Malta
   \and
     Jeremiah Horrocks Institute, University of Lancashire, Preston PR1 2HE, UK
    \and
     Instituto de Astrof\'isica de Andaluc\'ia (CSIC), Glorieta de la Astronom\'ia s/n, E-18008 Granada, Spain
    \and
    Department of Physics and Astronomy, University College London, London WC1E 6BT, UK
    \and
    Department of Physics, Engineering Physics, and Astronomy, Queen’s University, Kingston, ON, K7L 3N6, Canada
    \and
    Universit\'e C\^ote d’Azur, Observatoire de la C\^ote d’Azur, Laboratoire Lagrange, CNRS, Blvd de l’Observatoire, 06304 Nice, France   
    }
         
\date{Received December 22, 2025; accepted March 18, 2026}

  \abstract
    {There is now strong evidence that the Milky Way (MW) hosts a nuclear stellar disc (NSD). However, whether the NSD is purely axisymmetric or contains a nuclear bar remains unresolved. Since approximately $50\%$ of barred galaxies with MW-like mass in the local Universe host a nuclear bar, investigating whether the MW hosts one is of interest. We conduct a systematic analysis to identify robust kinematic diagnostics capable of determining whether the MW hosts a nuclear bar. Using N-body simulations, we explore the kinematic signatures indicative of a nuclear bar. Using the phase-space coordinates longitude $(\ell)$, latitude $(b)$, proper motions ($\mu_\ell$ and $\mu_{\rm b})$ and line-of-sight velocity $(v_{\rm los})$, we test various diagnostics assuming different nuclear bar orientations. We also evaluate how sample size, dust extinction and bar amplitude influence the efficacy of the diagnostics. We identify two independent kinematic diagnostics capable of revealing a nuclear bar in the MW: (1) the vertex deviation, $l_{\rm v}$, of the ($v_{\ell}-v_{\rm los}$) velocity ellipse; and (2) The asymmetry in the $\mu_{\ell}$ vs $\ell$ distribution. While both are impacted by the sample size and extinction, the vertex deviation proves more robust, especially when combining stars from multiple observational fields. We also assess the correlation between the line-of-sight velocity and the $h_3$ Gauss-Hermite moment ("skewness") of the line-of-sight velocity but find no clear distinction between an NSD and a nuclear bar based on this metric. Our results suggest that data from the current KMOS survey may allow a marginal detection of a nuclear bar using the vertex deviation method. A companion paper provides further validation and detailed analysis of this approach. Nonetheless, future surveys will provide the high quality data necessary to fully exploit the diagnostics outlined in this study.}


   \maketitle
%
\section{Introduction}\label{intro}

The central few hundred parsecs of the Milky Way (MW) contain a nuclear stellar disc (NSD): a dense, rapidly rotating stellar structure that dominates the gravitational potential at Galactocentric radii $30 < R < 300\pc$ \citep{Schultheis+2025}. The NSD has a radial scale length of $R \simeq 90\pc$, a vertical scale-height of $H \simeq 30\mhyphen45\pc$ and a stellar mass of $M_{\mathrm{NSD}}=10.5^{+1.1}_{-1.0}\times 10^8 M_{\odot}$ \citep{Launhardt+2002,Nishiyama+2013, Schodel+2014, Gallego-cano+2020, Sormani+2022}.

\begin{figure*}
    \includegraphics[width=\linewidth]{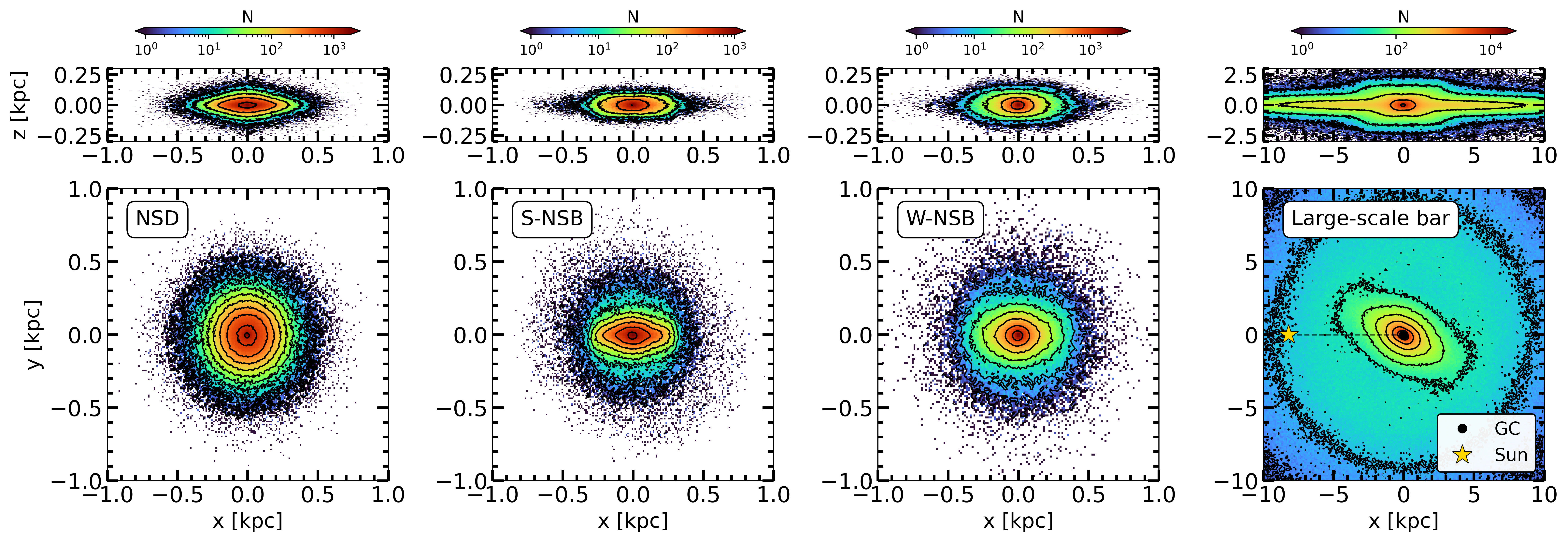}
    \caption{The face-on (bottom) and side-on (top) stellar density distribution of the models. Left to right: the axisymmetric NSD, isolated strong NSB, isolated weak NSB, and large-scale bar. The large-scale bar model is oriented at $-27\degree$ relative to the Solar-GC line (dotted line). The nuclear bar angle, $\alpha$, is a free parameter. In all cases, the Solar position is $(x,y,z) = (-8.2,0,0.012)$ kpc.}
    \label{fig:model_xy}
\end{figure*}

Whether the MW's NSD is purely axisymmetric remains an open question. Indeed, while a fair fraction of the current literature has treated the NSD as an axisymmetric structure \citep[e.g.,][]{Nogueras+2022}, some observational studies have previously suggested the existence of a nuclear bar in the MW: \citet{Alard+2001} used H- and K-band star counts from 2MASS to derive the projected stellar surface density in the bulge region. They reported a positive/negative longitude asymmetry which they interpreted as a signature of a nuclear bar. \citet{Rodriguez+2008} derived the mass distribution in the Galactic Centre by fitting the 2MASS star count map to a model including a disk, bulge and nuclear bar. Using this inferred gravitational potential, they performed hydrodynamical simulations to model the behaviour of the interstellar gas. The resulting gas flow patterns reproduced a characteristic parallelogram feature observed in the longitude-velocity diagram of molecular spectral lines, which they interpreted as evidence of a nuclear bar whose major axis is oriented at an angle of $\alpha \approx 60\degree\mhyphen75\degree$ relative to the Sun-Galactic Centre line. \citet{Nishiyama+2005} used red clump (RC) stars at a latitude of $b=+1\degree$ to trace the main bar orientation as a function of longitude ($-10.5\degree<\ell<10.5\degree$). They reported a change in the inclination angle for the Galactic bar at fields with $|\ell| < 4\degree$ (corresponding to a nuclear bar radius $R\sim570 \pc$), which they interpreted as the effect of a distinct inner bar structure. \citet{Gonzalez+2011} later confirmed this result using RCs from the VISTA Variables in the Via Lactea (VVV) survey to trace the orientation of the large-scale bar at latitudes $b = \pm 1\degree$. While the above results seem compelling, \citet{Gerhard+2012} employed N-body simulations to show that these observed asymmetries and the change in the slope of the star count profiles can be explained without invoking a nuclear bar. Instead, they demonstrate that the observed features arise naturally from the intrinsic density distribution of a single, barred, boxy bulge, which transitions from elongated to more axisymmetric within the inner few degrees. This transition causes an apparent flattening of the longitude profiles near $|\ell|\approx 4\degree$, similar to the one observed by \citet{Nishiyama+2005} and \citet{Gonzalez+2011}, mimicking the expected signature of a nuclear bar.

A key challenge in determining whether a nuclear bar is present in the MW stems from our edge-on perspective of the Galactic Centre, which is heavily obscured by non-uniformly distributed dust \citep[e.g.][]{Bally+1988, Gonzalez+2012, Schultheis+2014,Nogueras-lara+2021,Henshaw+2023}. Notably, most of the gas/dust ($\sim 3/4$) within $|\ell|<3.5\degree$ resides at positive longitudes \citep[e.g.,][]{Bally+1988,Henshaw+2023}, which complicates the use of star counts and photometric methods by producing an apparent deficit of stars at positive longitudes. This introduces significant biases in the results challenging the reliability of photometric analyses. 

An alternative approach to detecting a nuclear bar involves the use of stellar kinematics. \citet{Sormani+2022} developed a dynamical equilibrium model fitted to 1D kinematic data from the KMOS NSD survey \citep{Fritz+2021}. Although their results indicate that the model fits are consistent with an axisymmetric system, it remains uncertain whether the 1D velocity histograms employed in their analysis are adequate to uncover evidence of non-axisymmetry. Nonetheless, nuclear bars are relatively common features in MW–like galaxies, with roughly $50\%$ of barred galaxies with stellar masses above $\log(M_{\star}/M_{\odot})>10.5$ being double-barred \citep{Erwin+2024}. Therefore, the possibility that our Galaxy hosts a nuclear bar warrants careful consideration.

Despite the current small sample sizes of existing 6D kinematic datasets, the anticipated availability of large and high-quality datasets from upcoming surveys such as the KMOS VVVX-GalCen\footnote{\url{https://www.eso.org/sci/observing/PublicSurveys/KMOS-surveys-projects.html}} (Nogueras-Lara et al. in prep), the Galactic Bulge
Time Domain Survey with \textit{Roman} \citep{Terry+2023}, MOONS \citep{Gonzalez+2020} and JASMINE \citep{Kawata+2024} will enable the dissection of the NSD region in unprecedented detail, allowing for new insights into the formation and evolution of the Galactic Centre. The goal of this paper is to make testable kinematic predictions that can confirm or rule out the presence of a nuclear bar in the MW with future surveys.

This paper is organized as follows: Section~\ref{sec:model} describes the models employed in this study. Section~\ref{sec:vertex_deviation} and Section~\ref{sec:mu_l_assymetry} present the two main kinematic diagnostics for detecting a nuclear bar\footnote{We also tested Gauss-Hermite moments as a third diagnostic method, but found it unsuitable for detecting a nuclear bar (see Appendix~\ref{sec:gh_moment}).}: the vertex deviation and the asymmetry observed in the $\mu_{\ell}$ versus Galactic longitude, $\ell$, distribution. Section~\ref{sec:discussion} provides a discussion of the results and their implications. Lastly, Section~\ref{sec:future_work} briefly outlines prospects for future work.

\begin{figure*}
    \centering
    \includegraphics[width=.9\linewidth]{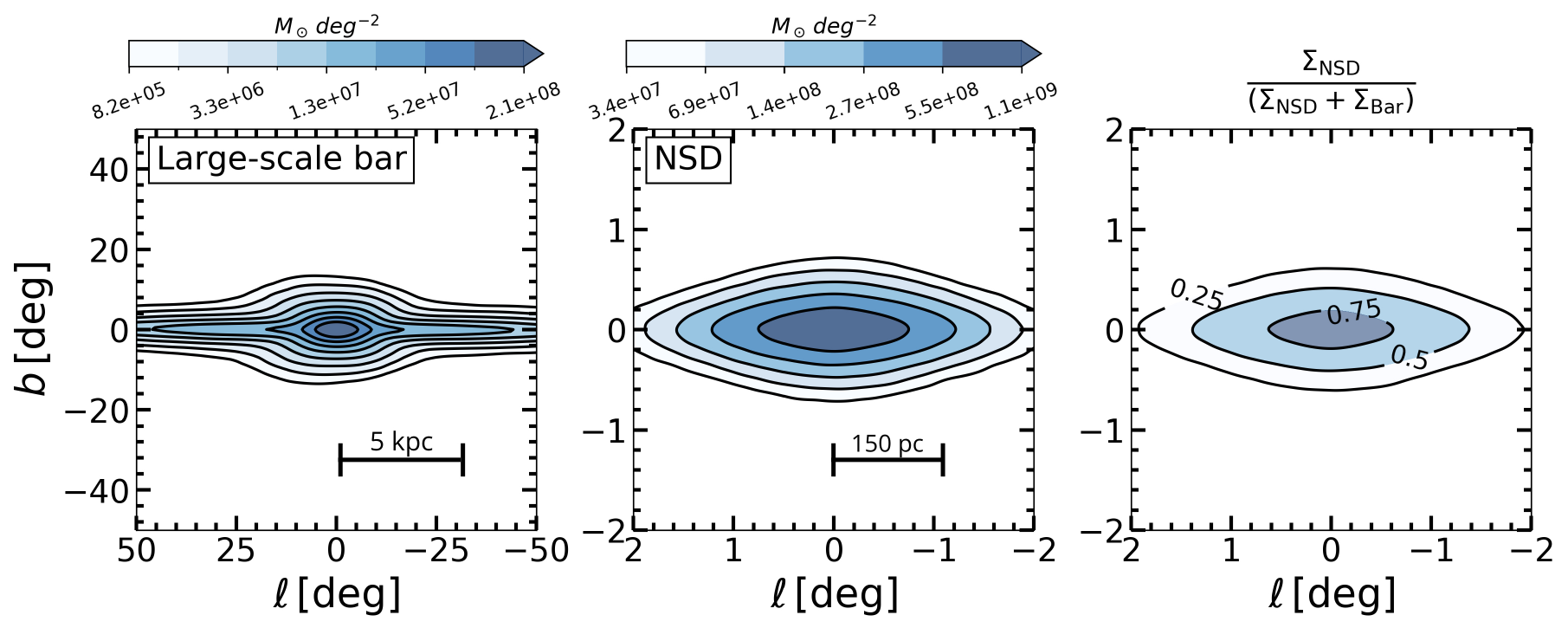}
    \caption{The surface density of the large-scale MW model with no nuclear component (left), the isolated NSD (middle), and the ratio
between the two (right). The ratio illustrates how prominent the MW’s NSD would be if we were to see it from the Sun (see Fig.~\ref{fig:model_xy}).}
    \label{fig:model_density}
\end{figure*}

\begin{figure}
    \centering
    \includegraphics[width=.90\linewidth]{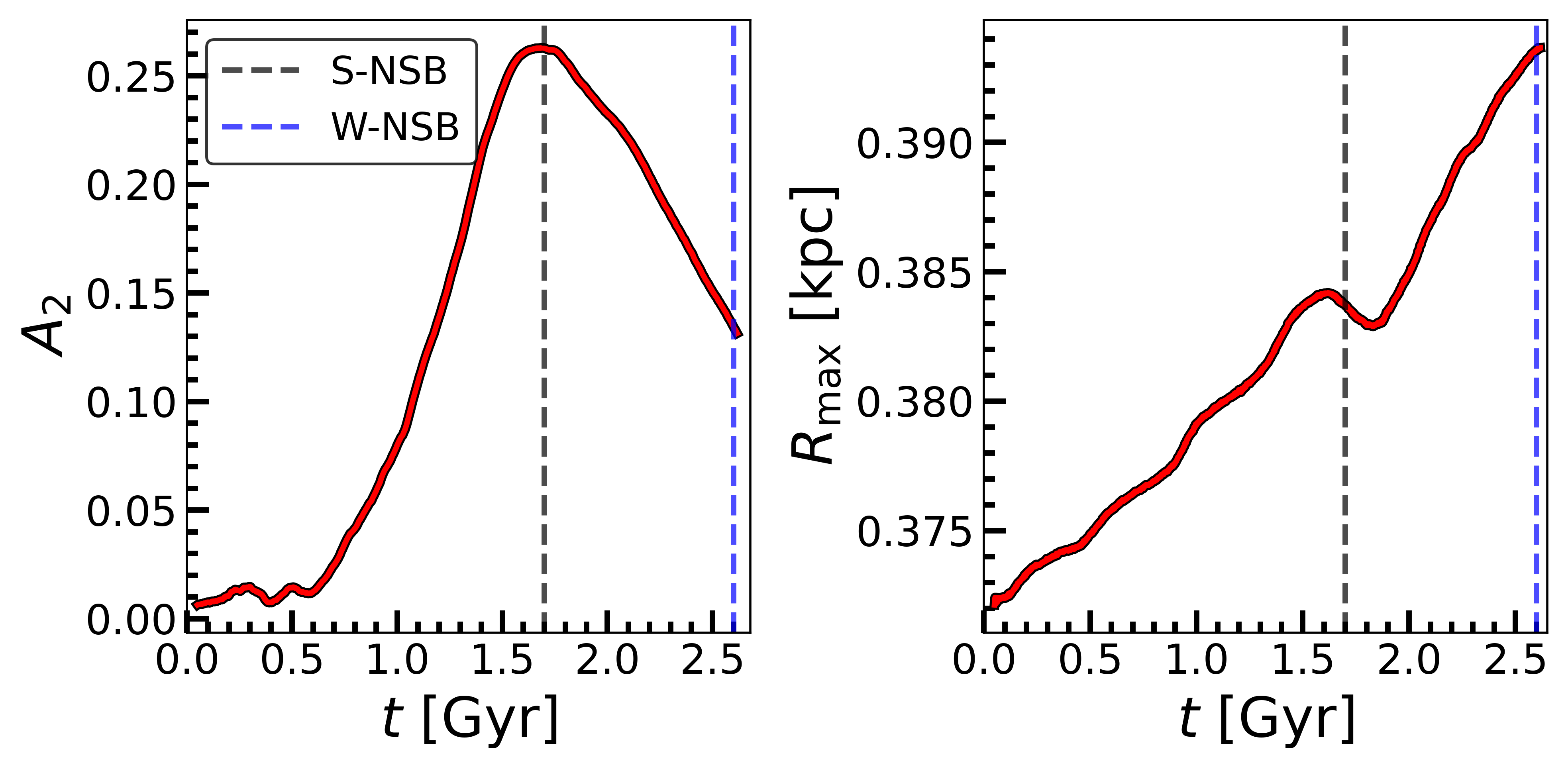}
    \caption{Evolution of the NSD model. \emph{Left:} the bar strength, measured by the Fourier $A_2$ parameter, as a function of time. $A_2$ is calculated considering all stars within the radius $R_{\rm max}$ that encloses $98\%$ of the model's mass.  \emph{Right:} the evolution of $R_{\rm max}$ with time. The initially stable NSD starts forming a bar at $t\approx0.5 \Gyr$, which reaches maximum strength at $t\approx1.7 \Gyr$. Black and blue dashed lines indicate the times at which the S-NSB and W-NSB models are extracted, respectively.}
    \label{fig:nsb_a2}
\end{figure}

\section{The models}\label{sec:model}
We consider three models of nuclear structure:
\begin{itemize}
\item \textbf{NSD}: an axisymmetric nuclear stellar disc (Sect.~\ref{sec:NSD})
\item \textbf{W-NSB}: a weak nuclear bar (Sect.~\ref{sec:NSB})
\item \textbf{S-NSB}: a strong nuclear bar (Sect.~\ref{sec:NSB})
\end{itemize}
In addition, we employ a large-scale MW model that includes the outer bar (Sect.~\ref{sec:MWmodel}). We construct the full models by embedding the nuclear components at the centre of the large-scale MW model, which includes an outer bar but no nuclear structure. Note that for the rest of the paper, when referring to the S-NSB and NSD models, these are always embedded in the large-scale bar. Otherwise, we refer to the "isolated" NSD or NSB models. Fig.~\ref{fig:model_xy} shows edge-on (top) and face-on (bottom) views of the models. Fig.~\ref{fig:model_density} shows the surface density in the plane of the sky of the large-scale MW model without a nuclear component (left), the surface density of the isolated NSD model (center), and the ratio between the two (right). To create the density panels, we rescale the large-scale bar model (see Sect.~\ref{sec:MWmodel} for details) to achieve an NSD/bar surface density ratio consistent with \citet{Sormani+2022}. The combined models are then constructed by randomly sampling stars from each component (nuclear and large-scale bar), with the relative sampling probability at each position $(\ell,b)$ determined by the surface density ratio shown in the right panel of Fig.~\ref{fig:model_density}. This ensures the correct levels of contamination from large-scale bar stars. When sampling from the large-scale bar component, stars are drawn from the original, unscaled model to preserve the correct kinematic properties. The large-scale MW model is described in more detail in Sect.~\ref{sec:MWmodel}.

We note that vertical proper motions ($\mu_b$) could in principle be used to reduce contamination from large-scale bar stars, as these exhibit systematically higher $|\mu_b|$ compared to the kinematically cooler NSD \citep[e.g.,][]{Shahzamanian+2022, Nogueras+2022}. However, as we demonstrate in Appendix~\ref{sec:vertical_motion}, applying a conservative cut of $|\mu_b| > 5\, \mathrm{mas\, yr^{-1}}$ reduces bar contamination by at most $\sim2\%$, which has negligible impact on our kinematic diagnostics. 

Lastly, although the large-scale MW model and the NSD/NSB models are each internally self-consistent, we do not evolve them together under their combined gravitational potential. Therefore, the combined models are not fully self-consistent. However, this has a negligible impact on the results given the very different spatial scales of the models.
\subsection{The axisymmetric NSD model} \label{sec:NSD}

We adopt the axisymmetric, self-consistent dynamical model of the NSD from \citet{Sormani+2022}\footnote{\url{https://github.com/GalacticDynamics-Oxford/Agama/blob/master/py/example\_mw\_nsd.py}}. This model was fitted to the line-of-sight and proper motion kinematic distributions obtained by combining the KMOS infrared (K-band) spectroscopic survey of \citet{Fritz+2021} with a preliminary version of the VIRAC2 proper motion catalogue \citep{Smith+2025}. The final fiducial NSD model has a total mass of $M_{\rm NSD}=9.7 \times 10^8 M_{\odot}$, an exponential radial scale length of $R_{\rm disc}=74 \pc$ and a vertical scale length of $H_{\rm disc}=26 \pc$.

\subsection{The strong and weak NSB models} \label{sec:NSB}

To obtain isolated nuclear bar models, we evolve the isolated NSD model of \citet{Sormani+2022} within the self-consistent MW model of \citet{Binney+2023}\footnote{\url{https://github.com/GalacticDynamics-Oxford/Agama/blob/master/py/example\_self\_consistent\_model\_mw.py}} using the N-body code \textsc{GADGET-4} \citep{Springel+2021}. We use softening lengths of 5, 50, and 25 pc for the NSD, dark matter, and main stellar disc components, respectively. We find that the initially axisymmetric NSD is mildly unstable, and starts forming a nuclear bar after evolving for $0.6\Gyr$. The left panel of Fig.~\ref{fig:nsb_a2}, illustrates the evolution of the $A_2$ bar amplitude, computed via the usual $m=2$ Fourier moment \citep[e.g.,][]{Li+2024}, alongside the radius containing $98\%$ of the NSD's mass (right panel). The output cadence is sufficiently fine that the $A_2$ profile is well-sampled across the full evolution; a running average is applied in Fig.~\ref{fig:nsb_a2} to reduce shot noise between successive outputs. The bar amplitude peaks at $t=1.7$ Gyr, after which it begins to weaken. We extract two snapshots from this simulation to construct our barred NSD models: the strong NSB (S-NSB) at $t=1.70 \Gyr$ (black dashed line) when the bar is at peak strength, and the weak NSB (W-NSB) at $t=2.60 \Gyr$  (blue dashed line) during the bar's declining phase.

\begin{figure*}
    \centering
    \includegraphics[width=.98\linewidth]{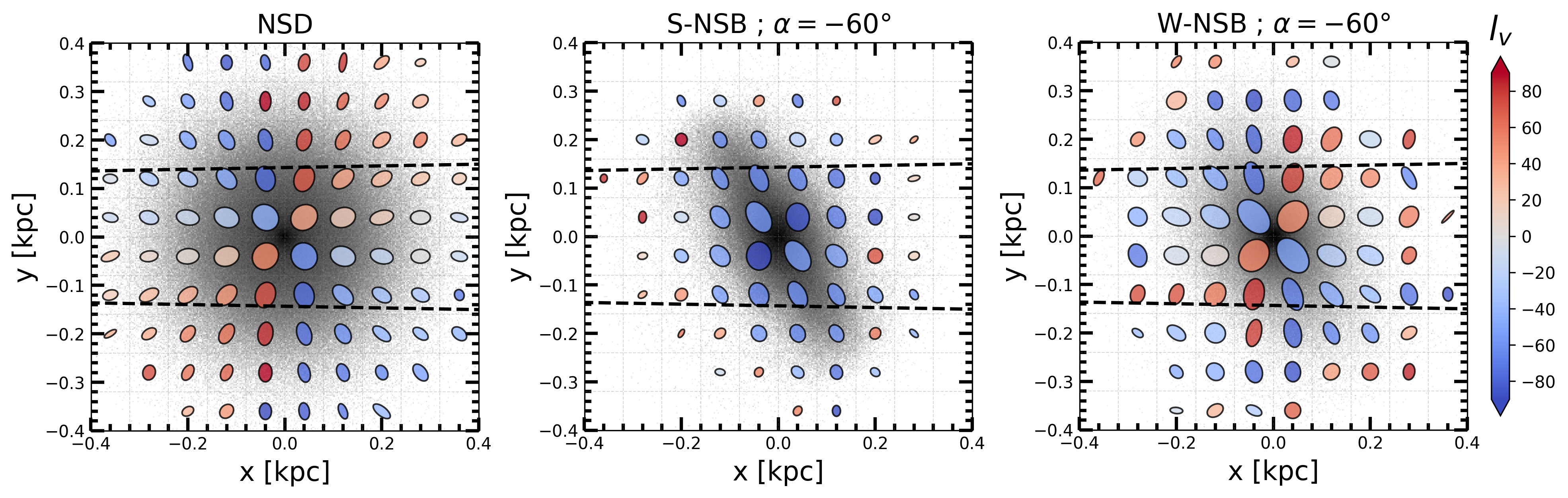}
    \caption{Plots in Galactocentric Cartesian coordinates of the isolated NSD (left), strong NSB (middle) and weak NSB (right) models oriented at $-60\degree$ relative to the Sun-GC line. Overlaid are the velocity ellipses coloured by the vertex deviation, \lv. We overlay an ellipse in each bin with $N>25$ stars. The dashed lines reflect a $2\degree$ field-of-view centred on the Solar position at $(x,y) = (-8.2, 0)\,\kpc$. The NSD shows radially oriented ellipses, while the S-NSB model ellipses show strong alignment with the bar's major axis. The alignment is weaker for the W-NSB model.}
    \label{fig:xy_ellipse}
\end{figure*}

\subsection{The large-scale MW model} \label{sec:MWmodel}

We employ a self-consistent N-body+Smooth Particle Hydrodynamic (SPH) model of a large-scale MW-like bar which corresponds to the model TG07v3 in \citet{Deg+2025}. Here, we provide only a brief overview.

The initial conditions consist of a bulgeless stellar disc with a S\'ersic surface density profile (S\'ersic index $n=1.75$), a gas disc (7\% of the stellar mass), and an NFW dark matter halo \citep{Navarro+1997} with mass $M_{\mathrm{h}}=1.17\times10^{12}\mathrm{M_\odot}$ and scale radius $R_{h}=19 \kpc$. The stellar disc has a mass of $M_d = 5.5 \times 10^{10} \, M_{\odot}$ and scale length $R_d = 0.43$ kpc (for the S\'ersic profile), while the gas disc has an exponential scale length of $R_g = 6.5$ kpc. The model is evolved using \textsc{Changa} \citep{Menon+2015} for $10 \Gyr$ using a base timestep of $\Delta t =0.5\Myr$, which provides sufficient temporal resolution that our results are insensitive to the choice of output cadence. Star formation occurs when gas density exceeds $0.1\, \mathrm{amu\, cm}^{-3}$ and temperature falls below $15{,}000\, \mathrm{K}$, with a star formation efficiency of 5\% per dynamical time and Supernova feedback via the \citet{Keller+2014} prescription.

The resulting model develops a bar that reaches a stable configuration by $\sim3.5$ Gyr and remains approximately constant thereafter. At $t=3$ Gyr, the bar has a semi-major radius of $R_{\mathrm bar}\sim4.25$ kpc and pattern speed of $\Omega\sim42\, \mathrm{km\,s^{-1}\,kpc^{-1}}$, consistent with measurements of the MW \citep[e.g.,][]{Wegg+2015, Portail+2017}. The model also reproduces the characteristic bimodal distribution of red clump stars observed along various lines of sight, mimicking the observed morphology of the MW's boxy/peanut bulge \citep{Gonzalez+2015}. We note that this model does not include a metal-poor spheroidal stellar component in the bulge; we assess the potential impact of such a population in Appendix~\ref{sec:metal_poor} and find it has negligible effect on our results.

To match the observed NSD/bar surface density ratio from \citet{Sormani+2022}, we rescale the model positions by $\alpha = 0.95$ and masses by $\gamma = 0.44$. This rescaling is essential to prevent the large-scale bar from dominating the surface density in the Galactic centre region. To maintain consistency with the laws of motion, we also rescale time by $\beta = 1.39$ and velocities by $\delta = 0.68$. The resulting scaled pattern speed is $\Omega_{\mathrm{scaled}} \approx 31\,\mathrm{km\,s^{-1}\,kpc^{-1}}$ and the bar semi-major axis is $R_{\mathrm{bar,scaled}} \approx 4.03\,\mathrm{kpc}$, both of which are consistent with observational constraints \citep[e.g.,][]{Minchev+2007, Portail+2015, Pontzen+2013, Sanders+2019, Bovy+2019, Chiba+2021, Clarke+2022}.

\section{Diagnostic 1: Vertex deviation}\label{sec:vertex_deviation}

\begin{figure*}
    \centering
    \includegraphics[width=\linewidth]{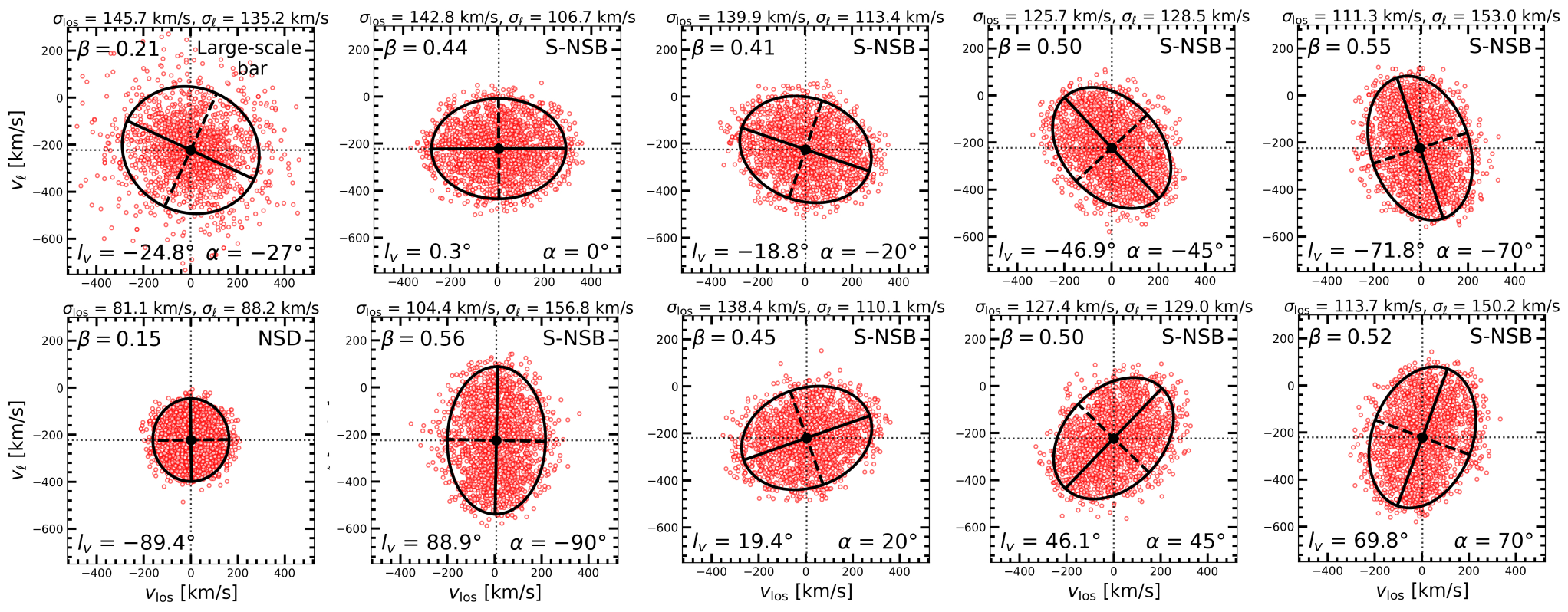}
    \caption{Velocity ellipses for the models, with the major and minor axes shown in solid and dashed lines, respectively. We uniformly sample $N=2,400$ stars from the region $|\ell|<0.9\degree$ and $|b|<0.25\degree$. In each panel we denote the vertex deviation (\lv), the anisotropy ($\beta$), the bar angle $\alpha$ (except for the NSD case), and the velocity dispersions in the line of sight and longitude directions. The top left panel shows the ellipse for the large-scale MW model with no nuclear component. As expected, the direction of highest dispersion is in the line of sight direction. The bottom left shows the ellipse for the axisymmetric NSD model. The other panels show the ellipses for the S-NSB model with various nuclear bar angles. In this case, the direction of highest dispersion depends on the nuclear bar angle.}
    \label{fig:all_ellipses}
\end{figure*}

\begin{figure}
    \includegraphics[width=\linewidth]{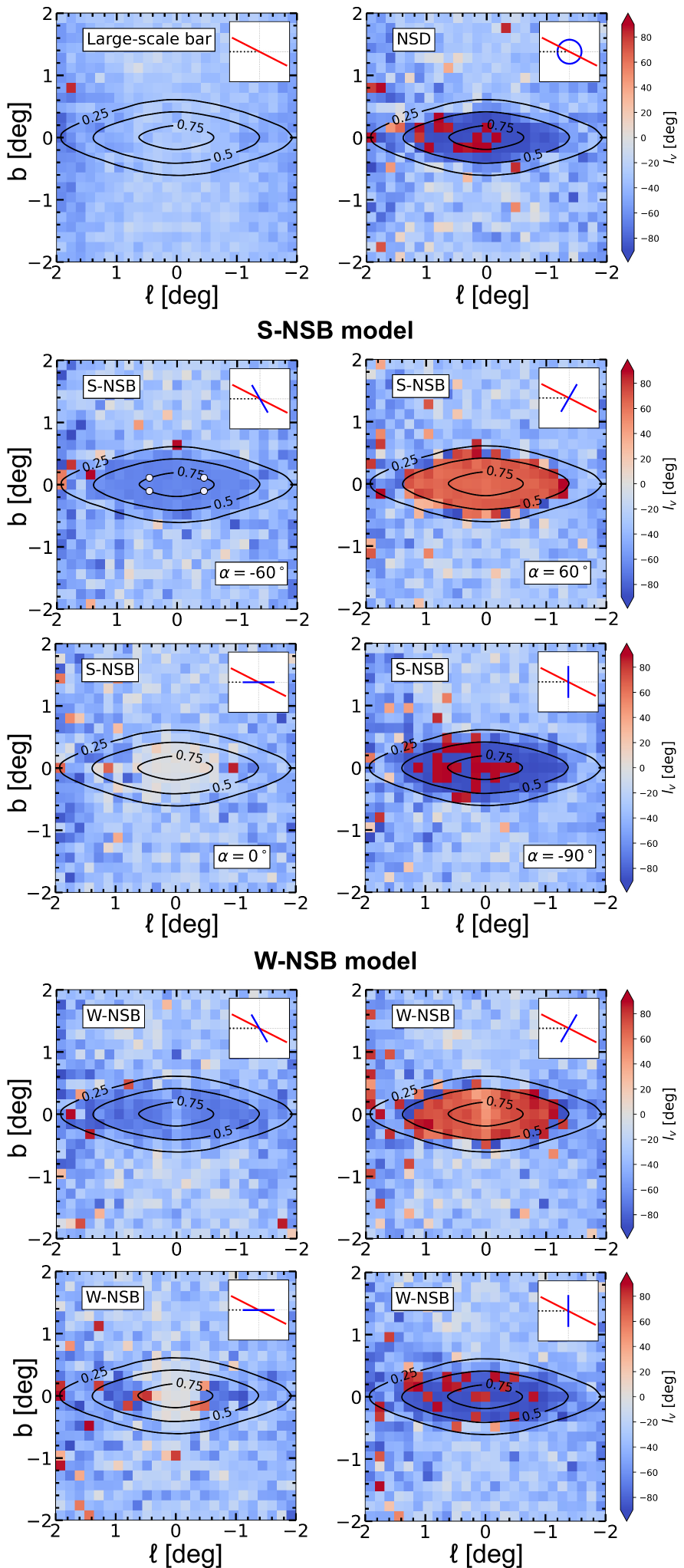}
    \caption{2D maps in $(\ell,b)$ space coloured by the vertex deviation, \lv. \emph{Top row}: Maps for the large-scale Milky Way model without a nuclear component (left) and with an axisymmetric NSD model (right); the NSD is indicated by the blue circle in the inset. \emph{Middle panels}: Maps for the S-NSB model with the nuclear bar at various orientations, $\alpha$. Insets show the geometric configuration of the outer bar (red line) and nuclear bar (blue line). White markers in the top-left panel indicate the fields used in the analysis of Sec.~\ref{sec:lv_n}. \emph{Bottom panels}: Same as above for the W-NSB model}
    \label{fig:vdev_maps}
\end{figure}

\begin{figure}
    \includegraphics[width=\linewidth]{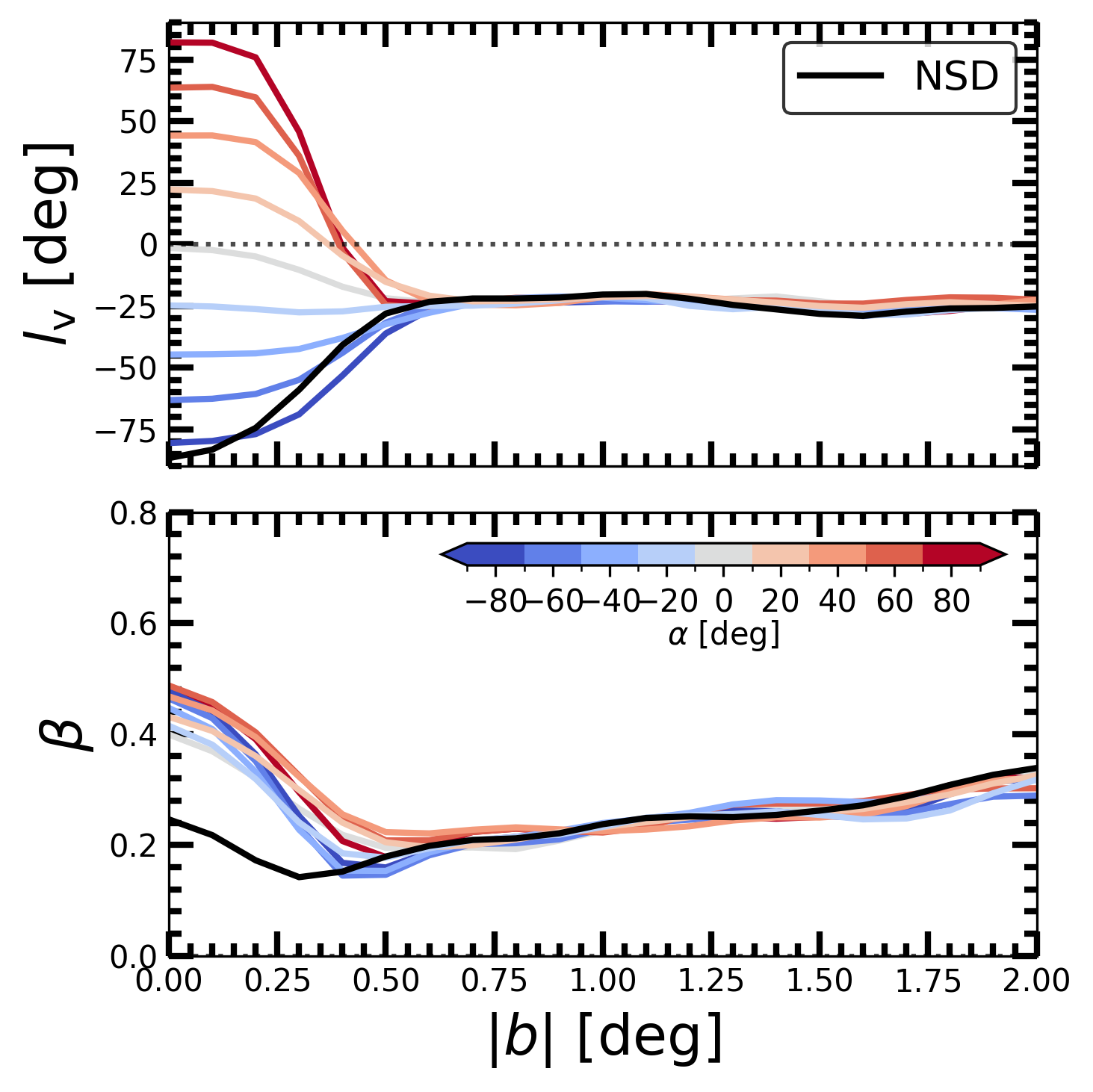}
    \caption{Vertex deviation, (\lv) and anisotropy, ($\beta$) as a function of Galactic latitude, $|b|$, for the NSD (black curves) and S-NSB model with varying values of the nuclear bar angle, $\alpha$ (coloured curves). As expected, the NSD has \lv\ $\approx 90\degree$ in the midplane. In contrast, the midplane values for the S-NSB model are dependent on $\alpha$. Across all model configurations, both \lv\ and $\beta$ converge to roughly similar values at $|b|\geq 0.5\degree$. This marks the latitude at which contamination from large-scale bar stars dominate the density distribution.}
    \label{fig:lv_lat}
\end{figure}

Given a set of stellar velocities, the velocity-dispersion tensor can be defined as
\begin{equation}
   \begin{aligned}
    \sigma^2_{ij} & = \langle (v_i-\langle v _i\rangle)(v_j-\langle v _j\rangle)\rangle \\ &=
    \langle v_i v_j \rangle-\langle v_i\rangle\langle v_j\rangle,
    \label{eqn:dispersion}
   \end{aligned}
\end{equation}
where $i$ and $j$ represent velocity components in the adopted orthogonal coordinate frame \citep{Binney+2008}. In the analysis in this section, we consider a 2D tensor where $i$ and $j$ can represent either the line-of-sight direction or the transverse direction parallel to the Galactic plane in the direction of increasing longitude $\ell$. We denote the velocities in these two direction as $v_{\rm los}$ and $v_\ell$ respectively. The dispersion tensor is symmetric and can therefore always be diagonalised. The orthogonal eigenvectors (corresponding to eigenvalues $\sigma_{\mathrm{max}}$ and $\sigma_{\mathrm{min}}$) define the principal axes of the velocity ellipse, with the major axis indicating the direction of maximum velocity dispersion (see Fig.~\ref{fig:schematic} and Appendix~\ref{sec:vertex_deviation_der}).

The orientation of the velocity ellipse in the ($v_{\rm los}$, $v_\ell$) plane can be characterised by the vertex deviation, \lv, which is given by:
\begin{equation}
    l_{\rm v}= \frac{1}{2}\arctan\left(\frac{2\sigma^{2}_{12}}{\sigma^{2}_{11}-\sigma^{2}_{22}}\right),
    \label{eqn:vertex_deviation}
\end{equation}
where $1$ and $2$ denote the $v_{\rm los}$ and $v_\ell$ directions respectively (see Appendix~\ref{sec:vertex_deviation_der}). By definition, \lv\ takes values $-90\degree \leq l_{\rm v} \leq 90\degree$. 

In an axisymmetric nuclear disc, since $\sigma^2_{12}\approx0$, the velocity ellipsoid of all stars along a given line of sight is expected to be approximately aligned with one of the coordinate axes (i.e., $l_{\rm v} \approx 0\degree$ or $l_{\rm v} \approx \pm 90\degree$) depending on the direction of highest velocity dispersion. For example, ellipses on opposite sides of the $x=0$ line in the left panel Fig.~\ref{fig:xy_ellipse} cancel out and their ``average'' orientation is aligned with the $x$ and $y$ axis. Note that strictly speaking this is valid only for an observer at infinite distance. For an observer at finite distance, the lines-of-sight that do not pass exactly through the Galactic centre (e.g. the dashed lines in Fig.~\ref{fig:xy_ellipse}) are not exactly parallel to the $x$ axis. However, the nuclear disc is small enough that this effect is negligible.

In the presence of a bar instead, stars preferentially stream along the bar's major axis, leading to a non-zero covariance $\sigma^2_{ij}$ and hence a non-vanishing \lv\, (see middle and right panels of Fig.~\ref{fig:xy_ellipse}). The vertex deviation has been successfully employed in both numerical \citep[e.g.,][]{Debattista+2019} and observational \citep[e.g.,][]{Simion+2021, Fernandez+2025} studies of the MW bulge to detect and characterise the Galactic bar.

The aspect ratio of the velocity ellipse is characterised by the anisotropy parameter:
\begin{equation}
    \beta= 1-\frac{\sigma^{2}_{\mathrm{min}}}{\sigma^{2}_{\mathrm{max}}},
    \label{eqn:beta}
\end{equation}
which takes values $0 \leq \beta \leq 1$. For $\beta=0$, the velocity distribution is isotropic (circular), and becomes increasingly elongated as $\beta \to 1$. One important limitation of using the vertex deviation is that \lv\ becomes increasingly unstable as $\beta \to 0$. In such cases, for a stellar population lacking a preferred direction along either coordinate axis (i.e., $\sigma_{\mathrm{min}} \approx \sigma_{\mathrm{max}}$), the vertex deviation becomes poorly defined and lacks statistical significance when assessed via bootstrapping (Eq.~\ref{eqn:vertex_deviation_error}).

\subsection{Error handling}\label{sec:error_handling}

To quantify the uncertainties, we employ a bootstrap resampling method. Starting with a sample of $N$ stars, we generate $B$ resampled datasets, each maintaining the original sample size. For each re-sampling, we compute the vertex deviation $l_{\rm v}^*$. This process yields a distribution of $l_{\rm v}^*$ values, from which the uncertainty is derived as:

\begin{equation}
    \sigma_{l_{\rm v}}= \sqrt{\frac{1}{B}\sum_{k=1}^{B}(l_{{\rm v},k}^*-l_{\rm v})^2},
    \label{eqn:vertex_deviation_error}
\end{equation}
where $l_{\rm v}$ is the estimate from the original sample, and $l_{\rm v,k}^*$ are the bootstrap samples. 

We also assess the impact of observational uncertainties in $v_{\ell}$ and $v_{\mathrm{los}}$ on the derived $l_{\mathrm{v}}$ measurements. Using typical KMOS uncertainties  ($\sim 10 \kms$ for $v_{\mathrm{los}}$ and $\sim 23 \kms$ for $v_\ell$ based on Fritz et al. 2021 and VIRAC2), we applied a Monte Carlo approach convolving model velocities with these errors. In agreement with \citet{Fernandez+2025}, we find that for a given sample size, $N$, these observational uncertainties are negligible ($\sigma_{\ell_{\mathrm{v}}}<5\degree$); therefore, all uncertainties reported use bootstrapping only.

\subsection{Vertex deviation in the models}\label{sec:model_vdev}

We explore whether the vertex deviation can serve as a diagnostic for the presence of a nuclear bar. Specifically, we investigate: (1) whether a nuclear bar produces an \lv\ signature which is distinct from that of an NSD, (2) the number of stars required to reliably measure \lv, and (3) the impact of dust extinction on the measurement. We first consider results from the strong nuclear bar model (S-NSB). The impact of bar amplitude will be assessed in Sec.~\ref{weak_nsb}.

To develop intuition for the spatial variation of \lv\ across the models, we present in Fig.~\ref{fig:xy_ellipse} face-on density maps of each nuclear model, subdivided into a $10\times10$ spatial grid. At each grid cell, we overlay the corresponding velocity ellipse measured in $(v_\ell, v_{\mathrm{los}})$ space. The dashed lines indicate a $2\degree$ field-of-view centred on the Galactic Centre as seen from the Solar position at $(x,y) = (-8.2, 0)\,\kpc$. In the isolated NSD case (left panel), the major axes of the velocity ellipses are oriented radially toward the Galactic Centre, reflecting the dominant contribution of line-of-sight velocity dispersion in this axisymmetric system. In contrast, for the strong NSB model (middle panel), the ellipses are strongly aligned with the major axis of the nuclear bar ($\alpha=-60^\circ$), tracing the streaming motions of stars on bar-supporting orbits. In the weak NSB case (right panel), the alignment of the ellipses along the bar major axis is substantially weaker, as expected from increased levels of symmetry.

In Fig.~\ref{fig:all_ellipses}, we present the velocity ellipses derived from our models. The ellipses are constructed using $N=2,400$ stars sampled uniformly from the region $|\ell|<0.9\degree$ and $|b|<0.25\degree$. Within each panel, key parameters are indicated: the vertex deviation \lv, the anisotropy $\beta$, the bar angle $\alpha$ (where applicable), and the velocity dispersions along the line-of-sight ($\sigma_{\mathrm{los}}$) and along the Galactic longitude ($\sigma_{\ell}$). The top-left panel illustrates the velocity ellipse for the large-scale bar model without a nuclear component. In this configuration, the velocity dispersion is highest along the line-of-sight direction, consistent with observations of the MW of metal rich $(\feh>0.30)$ stars tracing the large-scale bar \citep[e.g.,][]{Babusiaux+2010}. The bottom-left panel shows the velocity ellipse for the axisymmetric NSD model, which shows velocities only weakly longitudinally biased (also reflected by the low anisotropy $\beta=0.15$). As a result, the vertex deviation for the axisymmetric NSD model is $l_v\approx \pm90\degree$. The remaining panels display the velocity ellipses for the S-NSB model at various nuclear bar angles $\alpha$. Compared to the NSD model, the anisotropy is high ($\beta\approx0.50$) across all angles. As expected, the value of the vertex deviation, \lv\, roughly matches the nuclear bar angle, $\alpha$.

\begin{figure*}
    \centering
    \includegraphics[width=\linewidth]{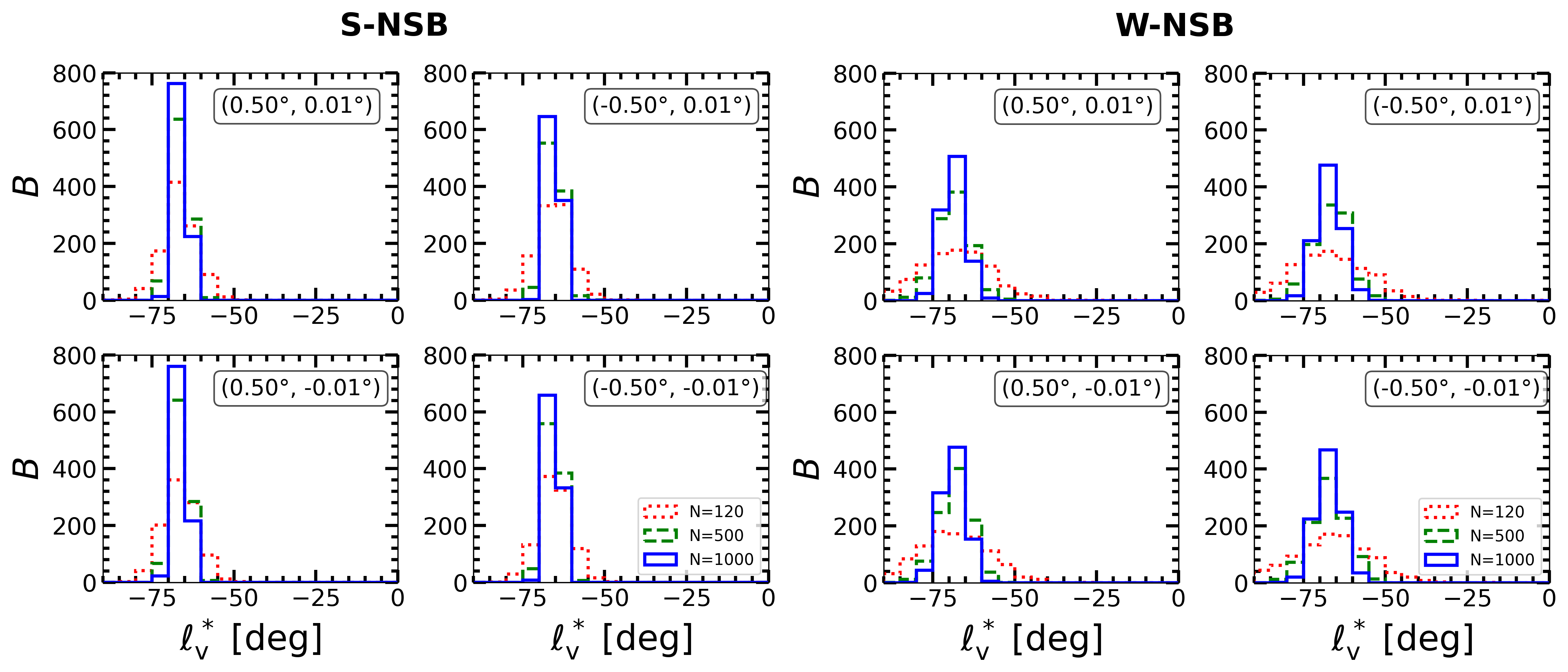}
    \caption{Using the S-NSB (left panels) and W-NSB (right panels) models with a nuclear bar angle $\alpha=-60\degree$, we sample stars from fields shown with $(\ell,b)$ coordinates indicated in each panel. We also indicate the position of each field (white markers) in the top left panel for the S-NSB model in Fig.~\ref{fig:vdev_maps}. For each field, we perform $B=1000$ bootstrap iterations using sample sizes of 120 (dotted red), 500 (dashed green), and 1000 (solid blue) stars. For the S-NSB model, the mean uncertainty (using Eq.~\ref{eqn:vertex_deviation_error}) decreases from $\pm 6.18\degree$ for 120 stars to $\pm 2.97\degree$ for 500 stars, and $\pm 2.01\degree$ for 1000 stars. We note that \lv\ varies marginally across fields, ranging from $-66.4\degree$ to $-62.8\degree$ (with a standard deviation of $\sigma_{\ell_v}\pm2.58\degree$). The uncertainty in \lv\ for the W-NSB case is $16.08\degree$,  $9.43\degree$ and  $6.52\degree$ for 120, 500 and 1000 stars per field respectively. Across the fields, we find that the median \lv\ value ranges between $-70.4\degree$ and $-60.1\degree$. This variation likely reflects both the intrinsic differences in stellar kinematics across fields and varying contamination from large-scale bar stars.}
    \label{fig:error_hist}
\end{figure*}

Fig.~\ref{fig:vdev_maps} presents 2D maps in $(\ell,b)$-space with bins coloured by \lv. Overlaid are the fractional density contours from the right panel of Fig.~\ref{fig:model_density} for reference. As a control, the top-left panel shows the large-scale bar model, which lacks a nuclear component. The inset shows the orientation of the large-scale bar (red line), which is $-27\degree$ relative to the Sun-GC line. Consistent with expectations, \lv\ is negative across the region, reflecting the outer bar's orientation \citep{Zhao+1994, Soto+2007, Babusiaux+2010, Debattista+2019, Simion+2021, Fernandez+2025}. 

The top-right panel of Fig.~\ref{fig:vdev_maps} shows the axisymmetric NSD model (indicated by the blue circle in the inset). As shown in Fig.~\ref{fig:all_ellipses}, the velocity ellipse of the axisymmetric NSD is characterised by low anisotropy with a slight bias towards higher dispersion in the Galactic longitude direction. From the view-point of the Sun, this results in $l_v\approx90\degree$ in the region $|\ell|\lesssim1\degree$ and $|b|\lesssim0.25\degree$. The middle and bottom set of panels in Fig.~\ref{fig:vdev_maps} present the results for the S-NSB and W-NSB models, respectively, with the nuclear bar oriented at various angles $\alpha$ relative to the Sun-GC line. The insets show the NSB orientation (blue line), while the outer bar remains fixed at $-27\degree$ in all cases. For both the S-NSB and W-NSB models, a nuclear bar at $\alpha = 60\degree$, produces positive values of \lv. More generally, nuclear bar orientations in the range $0\degree \leq \alpha \leq 90\degree$ yield positive \lv, whereas orientations in the range $-90\degree \leq \alpha \leq 0\degree$ produce negative \lv\ values. As we show later in Sec.~\ref{weak_nsb}, \lv\ varies smoothly with the NSB angle. In the limiting case of $\alpha = 90\degree$, the resulting signal is difficult to distinguish from that produced by an axisymmetric NSD.

Fig.~\ref{fig:lv_lat} shows the variation of \lv\ (top) and $\beta$ (bottom) with Galactic latitude $|b|$ for the NSD (black curve). The other curves show the variation for the S-NSB model with the nuclear bar at varying angles, $\alpha$. Here, we only consider stars satisfying $|\ell|<1\degree$. As expected, \lv\ for the NSD model is \lv$\approx 90\degree$ in the midplane, while this value changes for the S-NSB model depending on $\alpha$. We find that in the midplane  $\beta\approx 0.22$ for the NSD model, while it takes on larger values $(0.4<\beta<0.5)$ for the S-NSB model. In all cases, as $|b|$ increases, both \lv\ and $\beta$ converge towards a single value. This is due to increased contamination from stars associated with the large-scale bar. At $|b| \approx 0.5\degree$, the vertex deviation plateaus around \lv\ $\approx -27\degree$. This marks the transition point where \lv\ becomes fully dominated by the large-scale bar.

\subsection{The impact of sample size on $l_{\rm v}$ }\label{sec:lv_n}

We now explore the dependence of the precision of \lv\ on the sample size. In Fig.~\ref{fig:error_hist}, we show the distribution of $l_{\rm v}^*$ (defined in Eq.~\ref{eqn:vertex_deviation_error}) across four representative fields indicated by the white markers in the top-left S-NSB panel of Fig.~\ref{fig:vdev_maps}. The sky Galactic coordinates are denoted in each panel. The left and right group of panels show the results obtained using the S-NSB and W-NSB models, respectively. In all cases, the models have a nuclear bar angle $\alpha = -60\degree$; however, we have checked that the results are consistent across all $\alpha$. We perform the bootstrap re-sampling method with $B=1000$ iterations for each field (panel), considering sample sizes of $120$ (dotted red), $500$ (dashed green) and $1000$ (solid blue) stars per field. For the S-NSB case, the average uncertainty (across all panels) in \lv\ is $\sigma_{l_v}=\pm 6.18\degree$ for a sample of $120$ stars, which decreases to around $\sigma_{l_v}=\pm 2.97\degree$ for $500$ stars. We find that for the case of a strong nuclear bar (but see Sec.~\ref{weak_nsb} for the weak case), around $500$ stars in each field may suffice for a reliable measurement of \lv. On the other hand only $120$ stars per field may lead to spurious measurements. We also note that some variation in \lv\ is expected across the fields due to both the different levels of contamination from the large-scale bar, and due to the fields probing different kinematic regimes. Across the four fields in Fig.~\ref{fig:error_hist}, \lv\ ranges from $-66.4\degree$ to $-62.8\degree$ (with $\sigma_{\ell_v} = \pm1.13\degree$) for $N=1000$ (blue histogram). 

\begin{figure}
\centering
    \includegraphics[width=.76\linewidth]{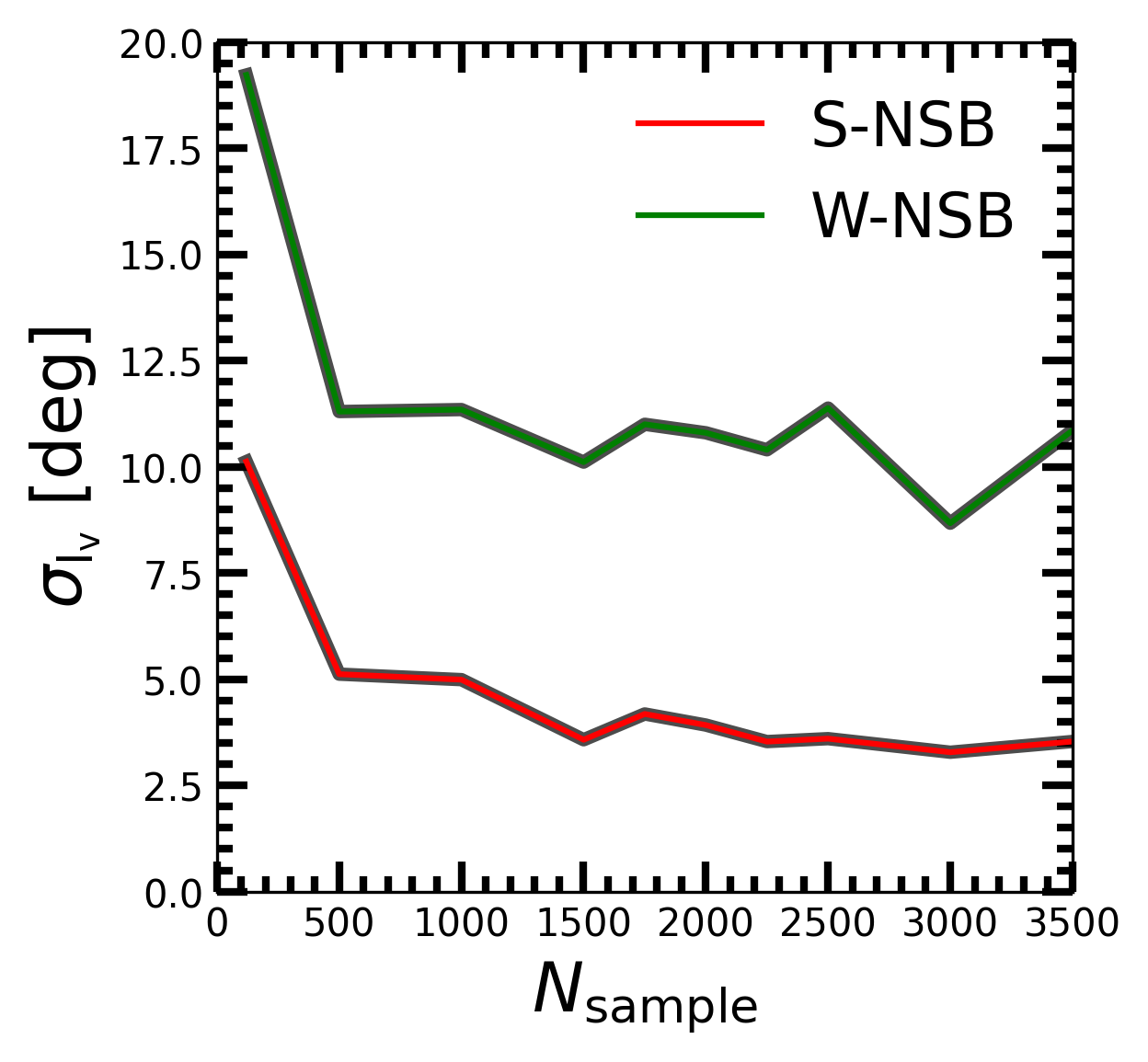}
    \caption{Uncertainty in the vertex deviation $\sigma_{l_{\rm v}}$ as a function of sample size, $N$. Stars are sampled from fields in Fig.~\ref{fig:error_hist}. We show the S-NSB (red) and W-NSB (green) cases, both with $\alpha=-60\degree$. The uncertainty decreases steeply up to $N=500$, then more gradually thereafter. The W-NSB case yields systematically higher uncertainties (by $\sim8\degree$) compared to S-NSB at all sample sizes due to its more isotropic velocity distribution (lower $\beta$).}
    \label{fig:error_plot}
\end{figure}

Combining stars from multiple fields to increase statistical significance of \lv\ is a potential strategy for further reducing the overall uncertainty. To illustrate this, Fig.~\ref{fig:error_plot} shows the uncertainty in \lv\ as a function of sample size, $N$, for stars sampled from the fields in Fig.~\ref{fig:error_hist}. The red and green curves show results for the S-NSB and W-NSB (discussed in Sec.~\ref{weak_nsb}) model, respectively. In both cases, the uncertainty decreases rapidly up to $N \approx 500$, then more gradually thereafter. We extend the analysis to $N=3500$, similar to the sample size in the current KMOS survey of \citep{Fritz+2021}.

\subsection{The impact of observational biases on $l_{\rm v}$ }\label{sec:lv_ext}

\begin{figure}
    \centering
    \includegraphics[width=\linewidth]{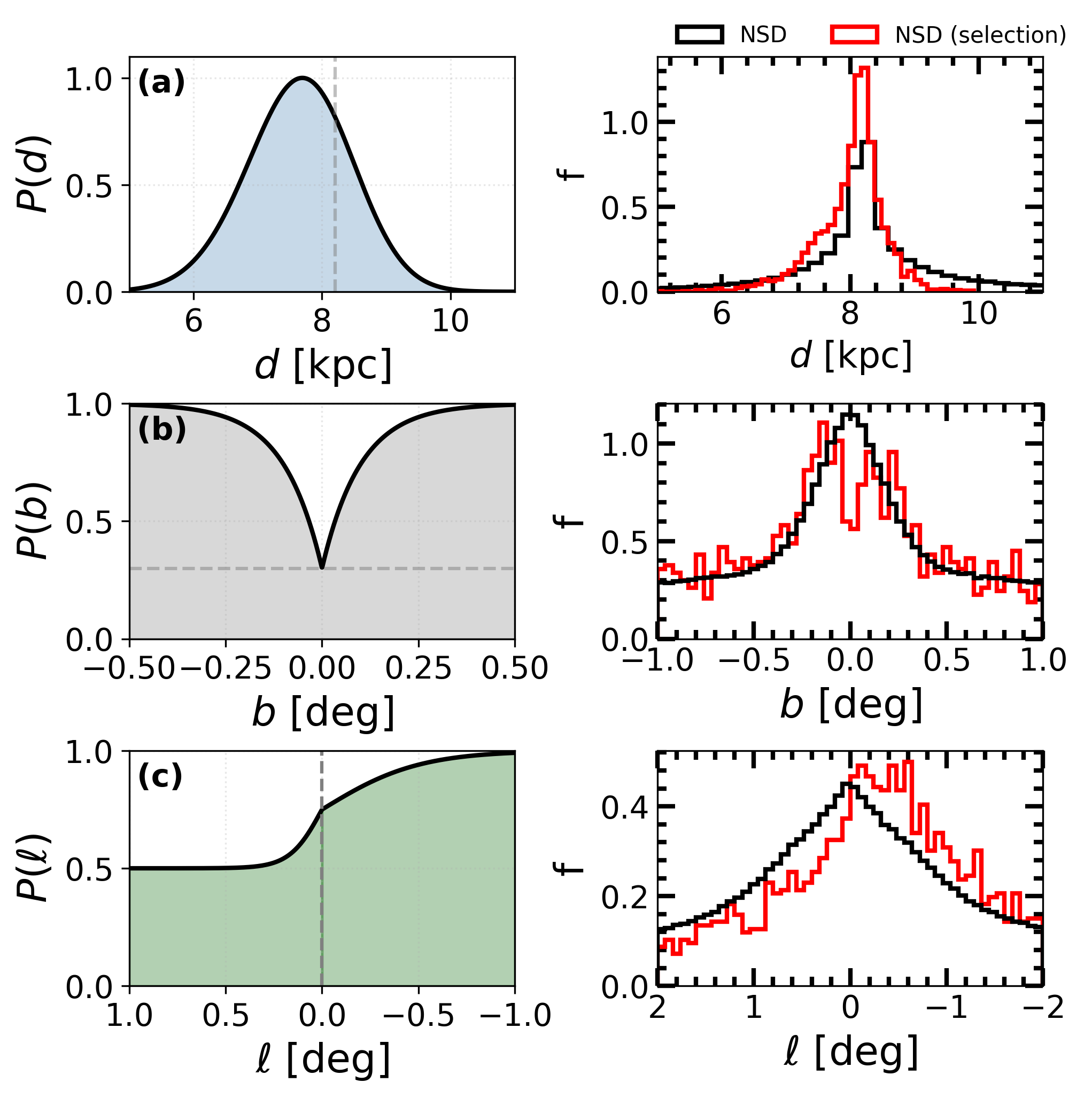}
    \caption{\emph{Left}: Observational selection functions applied to the models. \textbf{(a)} Distance-dependent detection probability favoring stars on the near side of the disc $(d<8.2\kpc)$. \textbf{(b)} Midplane extinction reducing completeness to 30\% at $b = 0°$. \textbf{(c)} Asymmetric longitude extinction with steeper decline at positive longitudes. The total probability is the product $P_{\rm total}(d, \ell, b) = P(d) \times P(b) \times P(\ell)$. \emph{Right}: Impact of selection on the NSD model's distance (top), latitude (middle), and longitude (bottom) distributions, showing the unbiased sample (black) versus the observed sample after applying selection effects (red).}
    \label{fig:selection}
\end{figure}

To assess whether observational selection biases could systematically affect our kinematic measurements and lead to a false detection of a nuclear bar, we applied a selection function to the axisymmetric NSD model and re-measured the kinematic parameters. Our selection function incorporates three physical effects: distance-dependent detectability (favouring the near side of the NSD), midplane extinction and crowding (suppressing detection near $b = 0\degree$), and asymmetric dust extinction (suppressing detection at $l > 0\degree$). Full details of the selection function are provided in Appendix~\ref{sec:selection}. The probability distribution functions for each of these are show in the left column of Fig.~\ref{fig:selection}.

The right column of Fig.~\ref{fig:selection} illustrates the impact of these selection effects on the stellar distributions. We compare the unbiased NSD model (black) to the "observed" sample after applying the selection function (red) for distance (top panel), Galactic latitude (middle panel), and Galactic longitude (bottom panel). The selected sample shows the expected biases: stars are preferentially drawn from the near side ($d < 8.2$~\kpc), higher latitudes ($|b| > 0.1\degree$), and negative longitudes where extinction is lower.

\begin{figure}
    \centering
    \includegraphics[width=\linewidth]{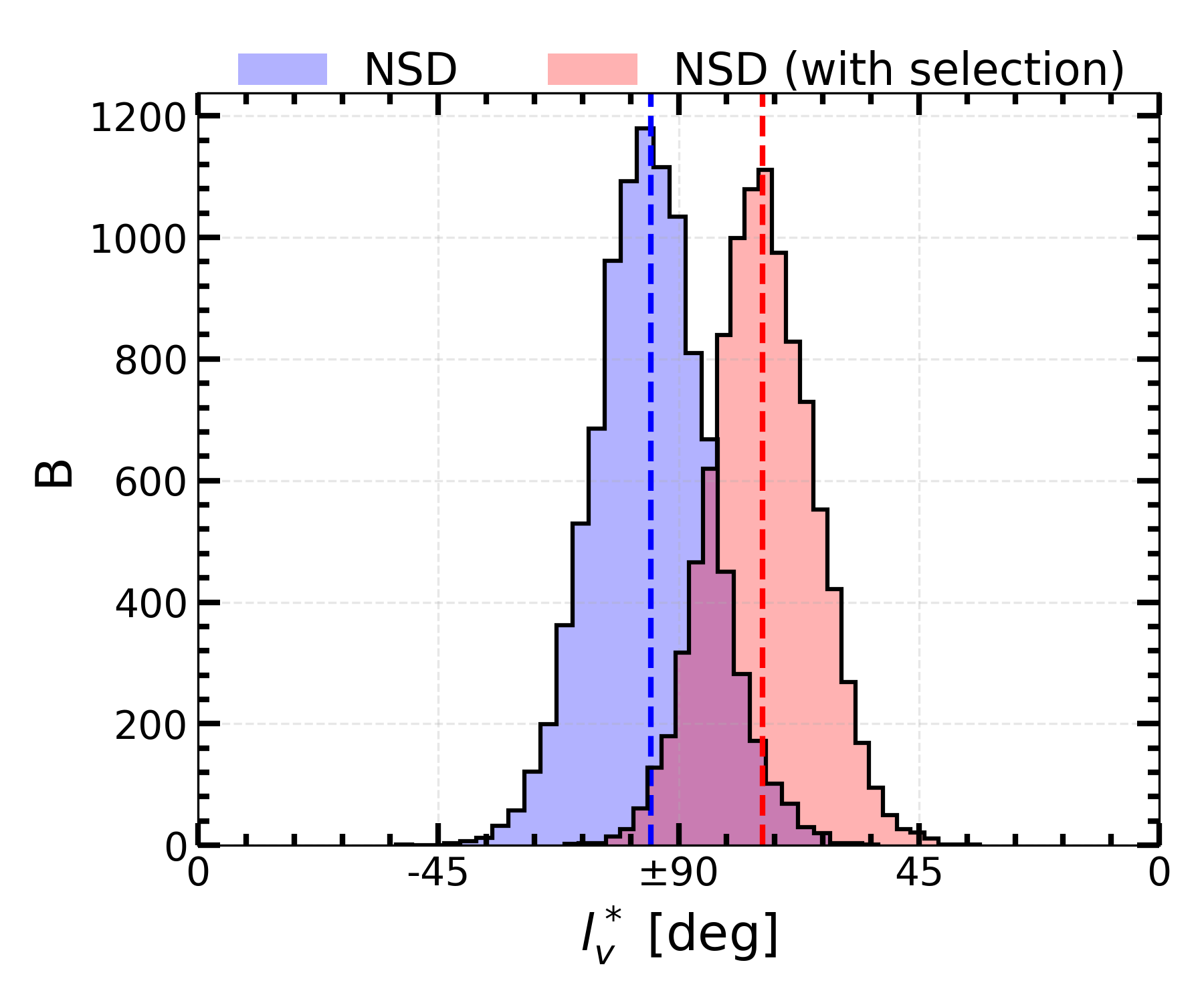}
    \caption{Bootstrapped distribution of $l^*_v$ for the axisymmetric NSD model without (blue) and with (red) the observational selection function applied. Each distribution results from 10,000 bootstrap iterations sampling $N=12,500$ stars from the region $|\ell|<0.9\degree$ and  $|b|<0.25\degree$. The unbiased NSD yields \lv$=(-87.4\pm 5.2)\degree$, consistent with an axisymmetric disc. However, applying realistic selection effects shifts the measurement to  \lv$=(82.2\pm 5.1)\degree$ (toward positive values), demonstrating that observational biases could mimic a nuclear bar with its near side at negative longitudes if not properly corrected.}
    \label{fig:lv_selection}
\end{figure}

Fig.~\ref{fig:lv_selection} shows the bootstrap distribution of the vertex deviation $l^*_v$ for the NSD model with (red) and without (blue) selection effects applied. Using a bootstrap approach, we resampled velocities for $N=12,500$ stars within the region ($|\ell|<0.9\degree$ and $|b|<0.25\degree$). For the NSD model without observational biases (blue histogram), we obtain a median value of \lv$=-87.4\pm 5.2\degree$. After applying the observational selection effects (red histogram), the vertex deviation measurement shifts to \lv$=82.2\pm 5.1\degree$.  

In general, we find that applying the selection function shifts the vertex deviation towards \textit{positive} values. To understand this effect, consider once more the left panel of Fig.~\ref{fig:xy_ellipse}. The combined probability function preferentially samples stars which are on the near side of the NSD and towards positive Galactic longitudes. This translates to stars which are in the bottom left quadrant $(x<0, y<0)$, where velocity ellipses with positive \lv\ dominate. Consequently, if extinction and observational biases are not carefully accounted for, this could lead to a spurious detection of a nuclear bar with its near side at negative Galactic longitudes.

We note that the distance bias component of our selection function (which preferentially samples foreground stars) has minimal impact on \lv\ when applied in isolation. This is because the distance bias affects stars on both sides of $\ell=0\degree$ symmetrically, and their velocity ellipse orientations effectively cancel when combined (see left panel of Fig.~\ref{fig:xy_ellipse}). The dominant systematic effect arises from the asymmetric longitude-dependent extinction, which can be partially corrected observationally through extinction mapping.

\subsection{A weak nuclear bar}\label{weak_nsb}

\begin{figure}
\centering
    \includegraphics[width=\linewidth]{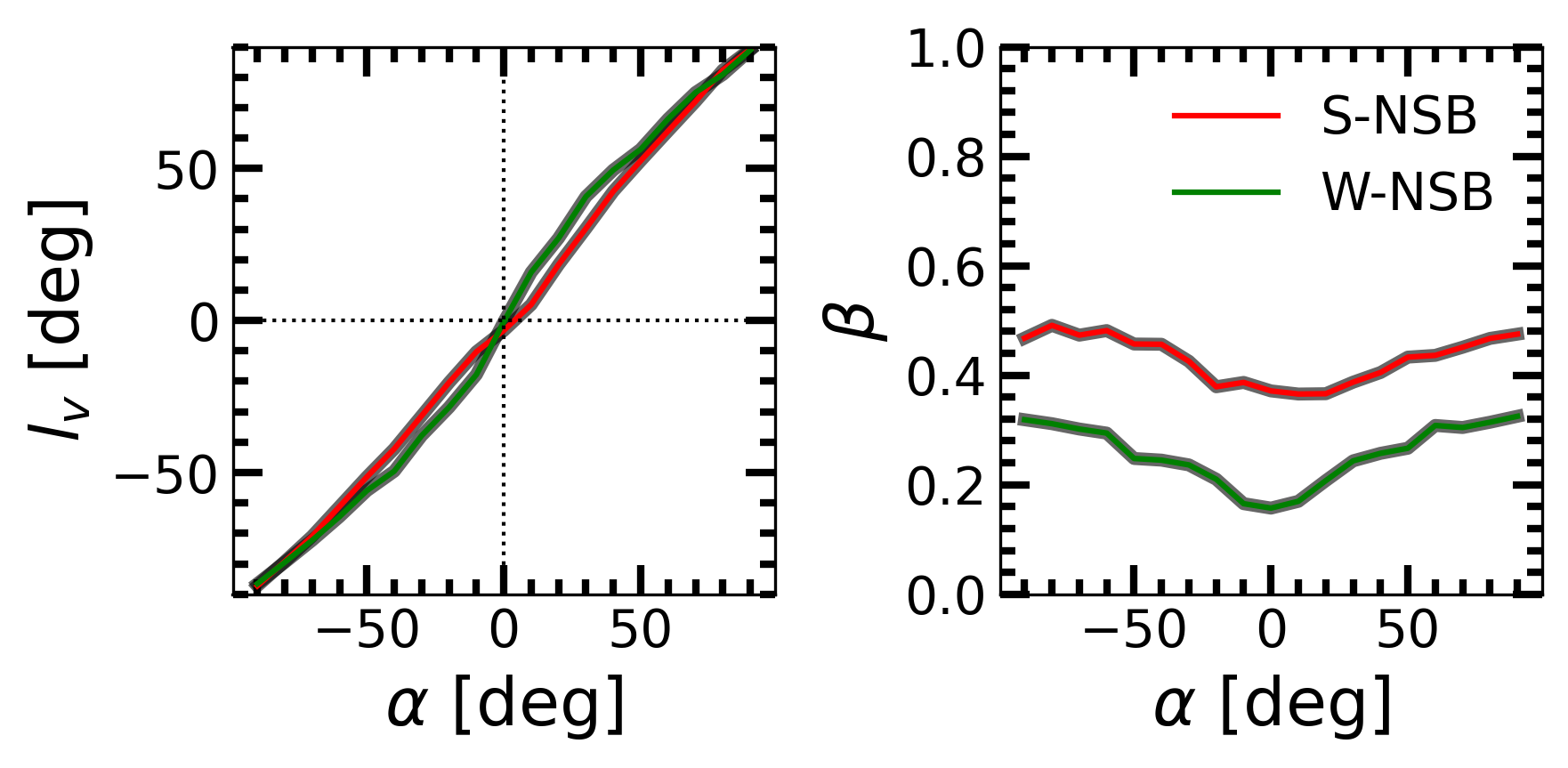}
    \caption{The vertex deviation (left) and anisotropy (right) of 1800 stars sampled from the region $|\ell|<0.9\degree$ and $|b|<0.25\degree$ as a function of nuclear bar angle, $\alpha$. Red curves show S-NSB results; green curves show W-NSB. While \lv\ remains roughly identical between strong and weak nuclear bars, $\beta$ decreases by approximately a factor of two for the weaker bar, approaching NSD-like values. This demonstrates that \lv\ is a robust orientation diagnostic but a blunt probe of bar amplitude, while $\beta$ may serve as a complementary bar amplitude indicator.}
    \label{fig:lv_a2}
\end{figure}

We now investigate how nuclear bar amplitude influences \lv\ and $\beta$ by repeating the vertex deviation analysis at $t \approx 2.60\,\mathrm{Gyr}$, when the nuclear bar has declined to approximately $65\%$ of its peak amplitude. Figure~\ref{fig:lv_a2} shows \lv\ (left) and $\beta$ (right) as a function of nuclear bar angle, $\alpha$, for the S-NSB (red) and W-NSB (green) models. The primary bar angle is fixed at $-27\degree$. In each case, we sample $1,800$ stars within the region $(|\ell|<0.9\degree, |b|<0.25\degree)$, which we combine to compute the kinematic properties.

Bar amplitude has minimal effect on \lv\ (left panel), consistent with the findings of \citet{Fernandez+2025}, who showed that \lv\ is a relatively blunt probe of non-axisymmetry. Specifically, in the presence of a non-axisymmetric structure, \lv\ is determined by the nuclear bar orientation $\alpha$, but it does not vary with the strength of the non-axisymmetry. In contrast, we find that $\beta$ decreases significantly for the weaker bar, by approximately a factor of two, resulting in values comparable to those of the NSD model $(\beta\approx0.15)$.

The more isotropic velocity distribution associated with the weaker bar (i.e., lower $\beta$) also leads to larger systematic uncertainties in \lv. In the right group of panels of Fig.~\ref{fig:error_hist}, we repeat the error analysis of Sec.~\ref{sec:lv_n} for the weak bar case. We find greater variation in the bootstrapped \lv\ values, with uncertainties of $16.08\degree$, $9.43\degree$, and $6.52\degree$ for samples of 120, 500, and 1000 stars per field, respectively. Across the fields, the median \lv\ value ranges between $-70.9\degree$ and $-45.7\degree$. This systematic increase in uncertainty is also evident in Fig.~\ref{fig:error_plot}. The red curve shows the uncertainty in \lv\ as a function of sample size when combining stars across the central fields. Compared to the strong bar case, we find a systematic increase in the uncertainty by approximately $8\degree$ across all values of $N_{\rm sample}$.

\subsection{An orthogonal nuclear disc}\label{sec:orthogonal}

\begin{figure}
    \centering
    \includegraphics[width=.8\linewidth]{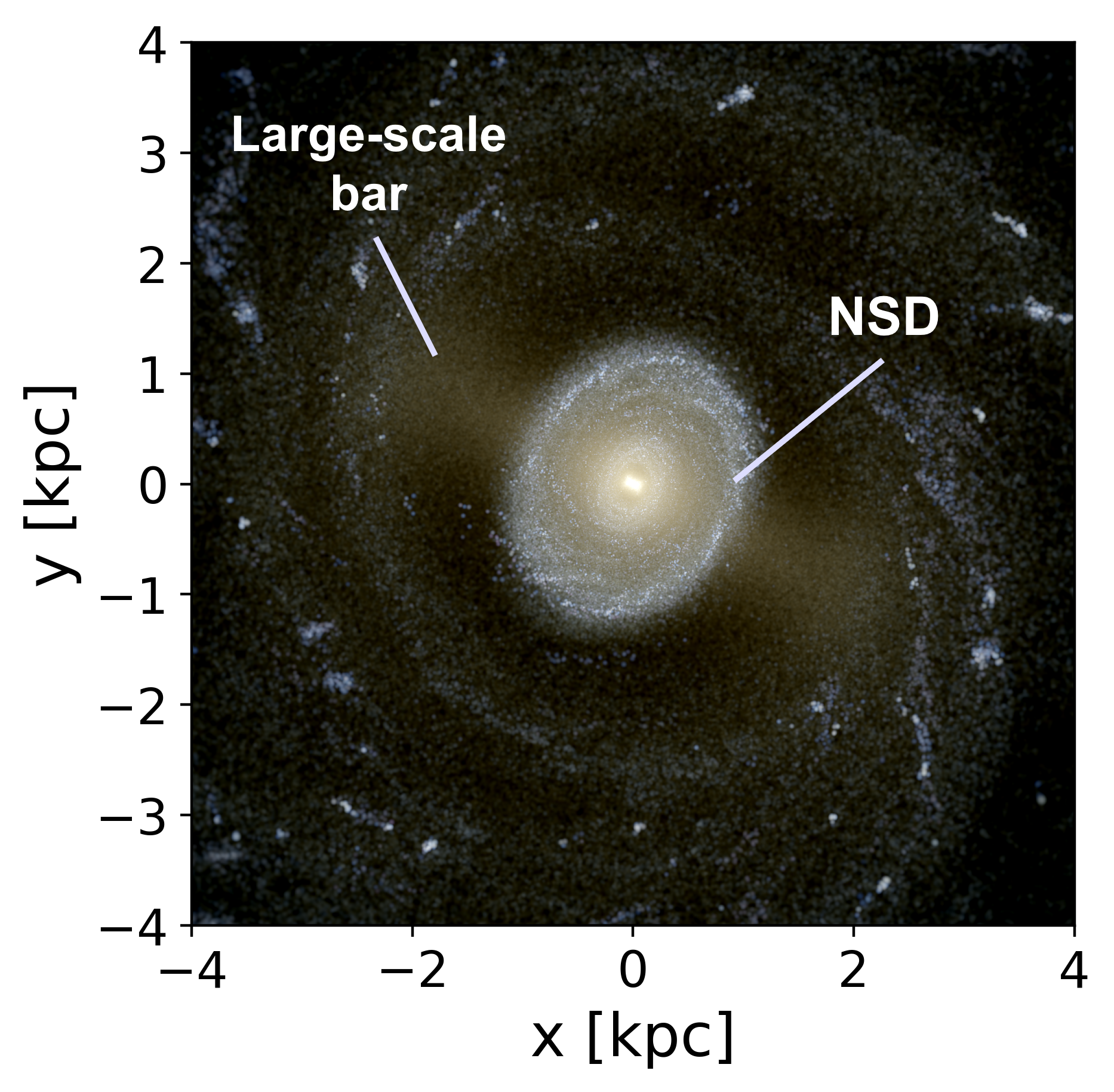}
    \caption{Synthetic RGB image produced with \textsc{pynbody} \citep{Pontzen+2013} of the stellar disc of the N-body+SPH model. We show the model face-on with the main bar aligned roughly $-27\degree$ to the Sun-GC line with the Sun located at $(x,y)=(-8.2,0)\kpc$. The central region of this model hosts an orthogonal sub $\sim1\kpc$ NSD with its near side toward negative Galactic longitude. This configuration tests whether orthogonal NSDs can be distinguished from true nuclear bars using kinematic diagnostics.}
    \label{fig:sph_render}
\end{figure}

\begin{figure}
\centering
    \includegraphics[width=\linewidth]{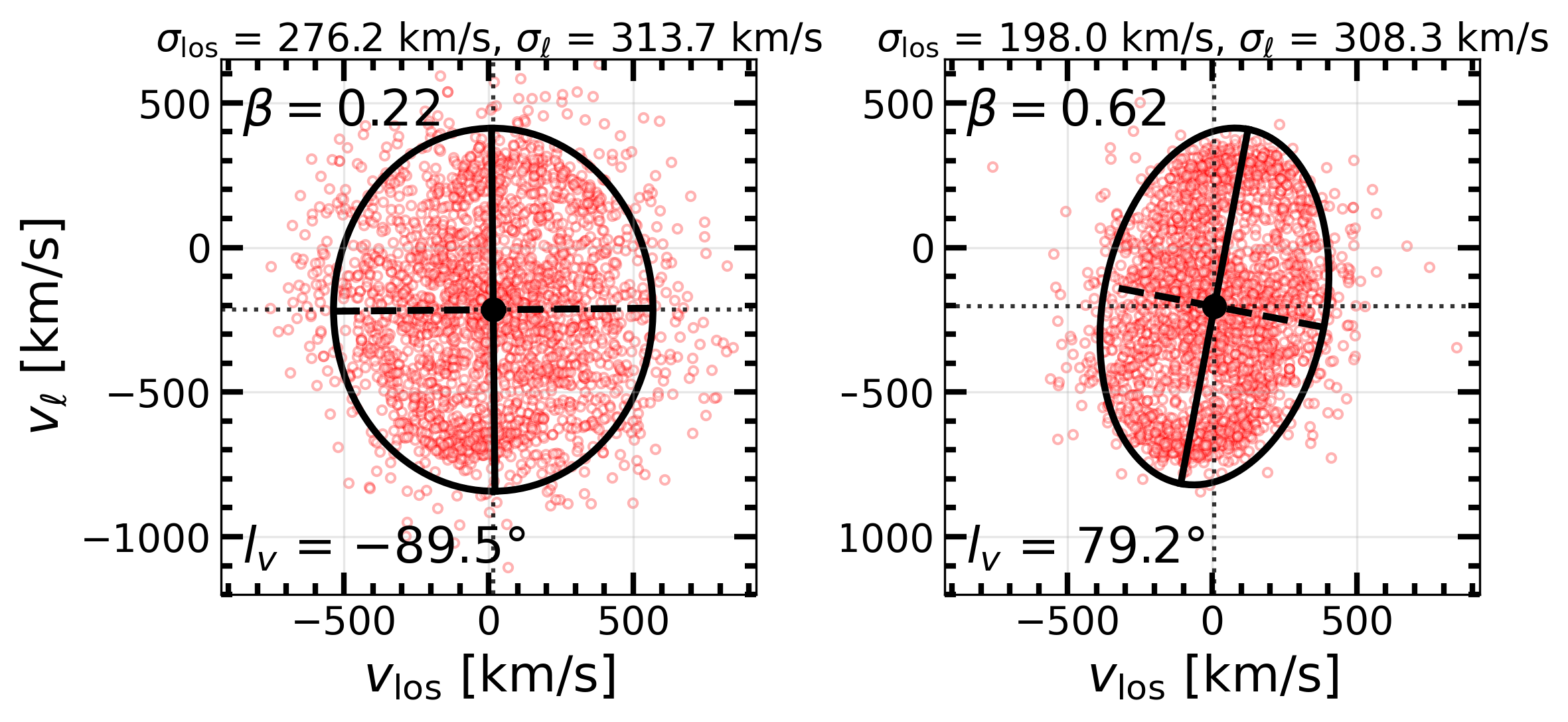}
    \caption{Velocity ellipses for the quasi-axisymmetric orthogonal NSD model, constructed from stars in the region $|\ell|<0.9\degree$ and $|b|<0.25\degree$. \emph{Left}: All stars show \lv$\approx -89.5\degree$ and low anisotropy ($\beta=0.22$), appearing nearly axisymmetric due to contamination from older large-scale bar stars. \emph{Right}: Selecting younger stars ($\tau_{\star}<5\,\mathrm{Gyr}$) which predominantly trace the late-forming NSD yields \lv$\approx 79.2\degree$ with higher anisotropy $\beta = 0.62$. This demonstrates that an orthogonal NSD can produce positive \lv, similar to a nuclear bar with its near side at negative longitudes.}
    \label{fig:sph_tilt}
\end{figure}

\begin{figure*}
    \includegraphics[width=\linewidth]{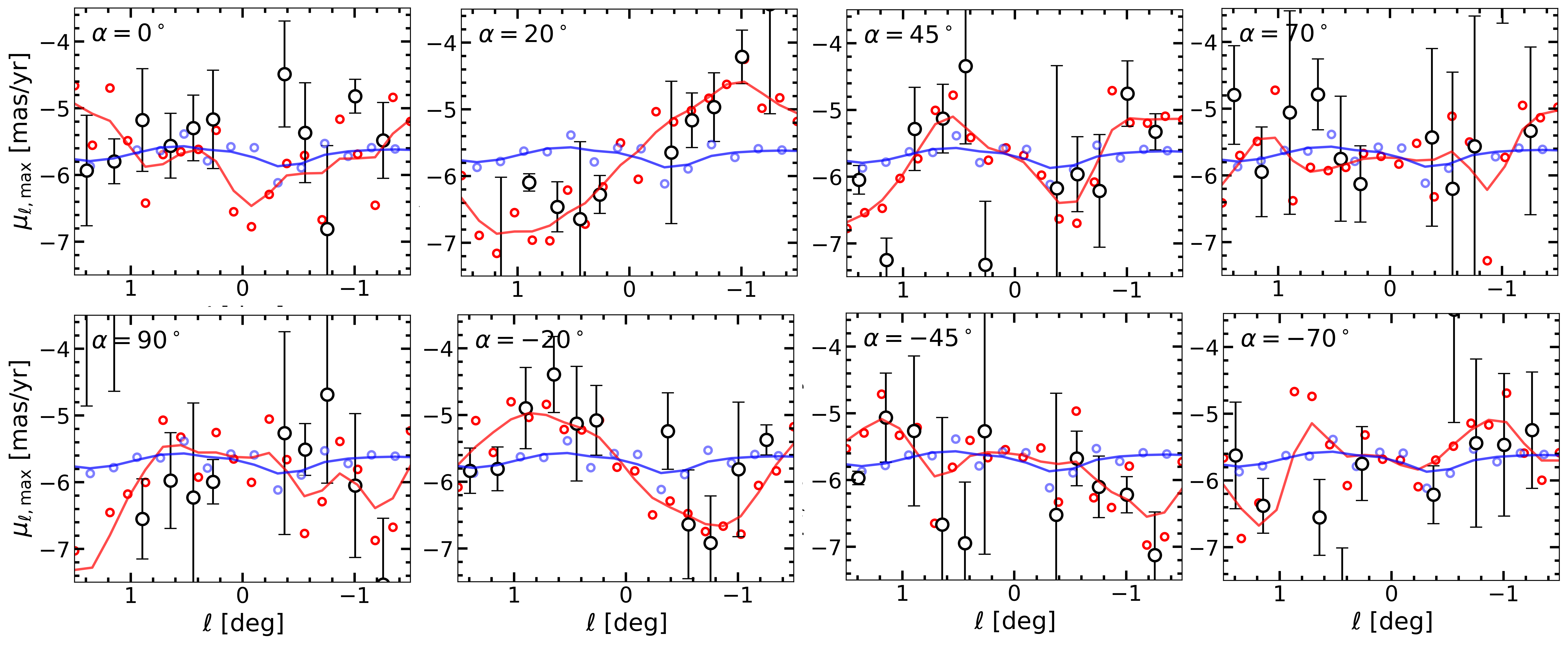}
    \caption{Distribution of $\mu_{\ell,max}$ versus $\ell$ distribution for the S-NSB model (red markers and smoothed fit) compared to the axisymmetric NSD (blue).  Each panel shows a different nuclear bar angle, $\alpha$. For small angles $|\alpha| \lesssim 20\degree$, the NSB produces a "step" feature in the central region ($|\ell|<1\degree$) that clearly deviates from the flat NSD distribution. This asymmetry weakens substantially as $|\alpha| \gtrsim$ increases, and becomes indistinguishable from the NSD at $\alpha = 0\degree$ or $\pm 90\degree$, limiting this diagnostic's applicability to a narrow range of nuclear bar orientations. Black open circles show measurements from individual midplane fields with $N=1500$ stars each from the S-NSB model. Uncertainties are estimated via bootstrap resampling.}
    \label{fig:mul_vs_l}
\end{figure*}

External galaxies frequently host elliptical NSDs \citep[e.g.,][]{Bittner+2020}. In barred galaxies, NSDs are predominantly supported by the $x_2$ orbital family, with the NSD's major axis oriented orthogonal to the primary bar \citep{Cole+2014}. Since the \citet{Sormani+2022} NSD model assumes axisymmetry, it cannot capture this orthogonal configuration. We therefore adopt a self-consistent N-body+SPH model (see Appendix~\ref{sec:sph_model} for details) to investigate the kinematic signature of an elliptical, orthogonally oriented NSD. We show a synthetic RGB rendering of the model in Fig.~\ref{fig:sph_render}, which features a primary bar with radius $R \approx 3\,\kpc$ and a $\sim 1\,\kpc$-scale NSD. While not MW-scaled, the large NSD allows us assess how an orthogonal configuration impacts the \lv\ signal.

As before, the primary bar is oriented at $-27\degree$ to the Sun--GC line, with its near side at positive longitudes. The orthogonal NSD has its near side toward \textit{negative} longitudes, implying it should exhibit positive \lv\ if the large-scale bar shows negative \lv. We  sample $2,400$ stars within the galactocentric radius $R < 1\,\kpc$ which also satisfy ($|\ell|<1\degree$, $|b|<0.3\degree$). The left panel of Fig.~\ref{fig:sph_tilt} shows the velocity ellipse for \textit{all} stars satisfying the selection criterion. The resulting \lv\ and $\beta$ values which are consistent with the axisymmetric NSD case (see Fig.~\ref{fig:all_ellipses}). Since the NSD in this model formed late, we re-sample the model again (with $N=2,400$) only considering stars with ages $\tau_{\star} < 5\,\mathrm{Gyr}$. This effectively reduces contamination from the large-scale bar. For this younger population, $\beta$ increases by approximately a factor of $\sim 3$. We also observe that \lv\ shifts away from the coordinate axis, towards positive values.

Therefore, an orthogonally oriented NSD in the MW would produce positive \lv\ signal, mimicking the kinematic signature of a nuclear bar viewed nearly side-on with its semi-major axis at negative longitude (see Fig.~\ref{fig:lv_a2}). This degeneracy implies that weak positive \lv\ in the Galactic Centre is consistent with either scenario and does not exclude the presence of a mildly elliptical NSD.

\section{Diagnostic 2: Asymmetry in the $\mu_{\ell}$ - $\ell$ plane}\label{sec:mu_l_assymetry}

The  $(\mu_{\ell}$,$\ell)$ plane provides a potential diagnostic for identifying an edge-on nuclear bar in the Galactic Centre. In an axisymmetric disc, the distribution of stars in the $(\mu_{\ell},\ell)$ plane is symmetric about the $\ell=0\degree$ line. Conversely, the presence of a nuclear bar can induce, depending on the orientation of the major axis, a measurable deviation from this symmetry. 

To capture these asymmetries, we first bin stars by their Galactic longitude $\ell$. Within each bin, a 1D histogram of $\mu_{\ell}$ is constructed and smoothed using Gaussian kernel density estimation (KDE). We identify the peak of each histogram, denoted as $\mu_{\rm \ell,max}$. Then we plot the relationship between $\mu_{\rm \ell,max}$ and $\ell$.

Fig.~\ref{fig:mul_vs_l} shows $\mu_{\ell,max}$ versus $\ell$ for the S-NSB model (hollow red markers) with various nuclear bar angles, $\alpha$. A smoothed fit to these points is shown by the red line. Blue markers and lines depict the corresponding distributions for the axisymmetric NSD case. The leftmost panels show the cases for $\alpha=0\degree$ (top) and $\alpha=90\degree$ (bottom), where the $\mu_{\ell,max}-\ell$ relationship is roughly consistent with that of the axisymmetric NSD. In contrast, for smaller $\alpha$ (second column), the distribution shows a "step" feature in $\mu_{\ell,\text{max}}$ which is not visible in the NSD case. As $\alpha$ increases (third/fourth columns) the prominence of the step features decreases quickly, eventually becoming flat. 

\begin{figure}
    \centering
    \includegraphics[width=0.8\linewidth]{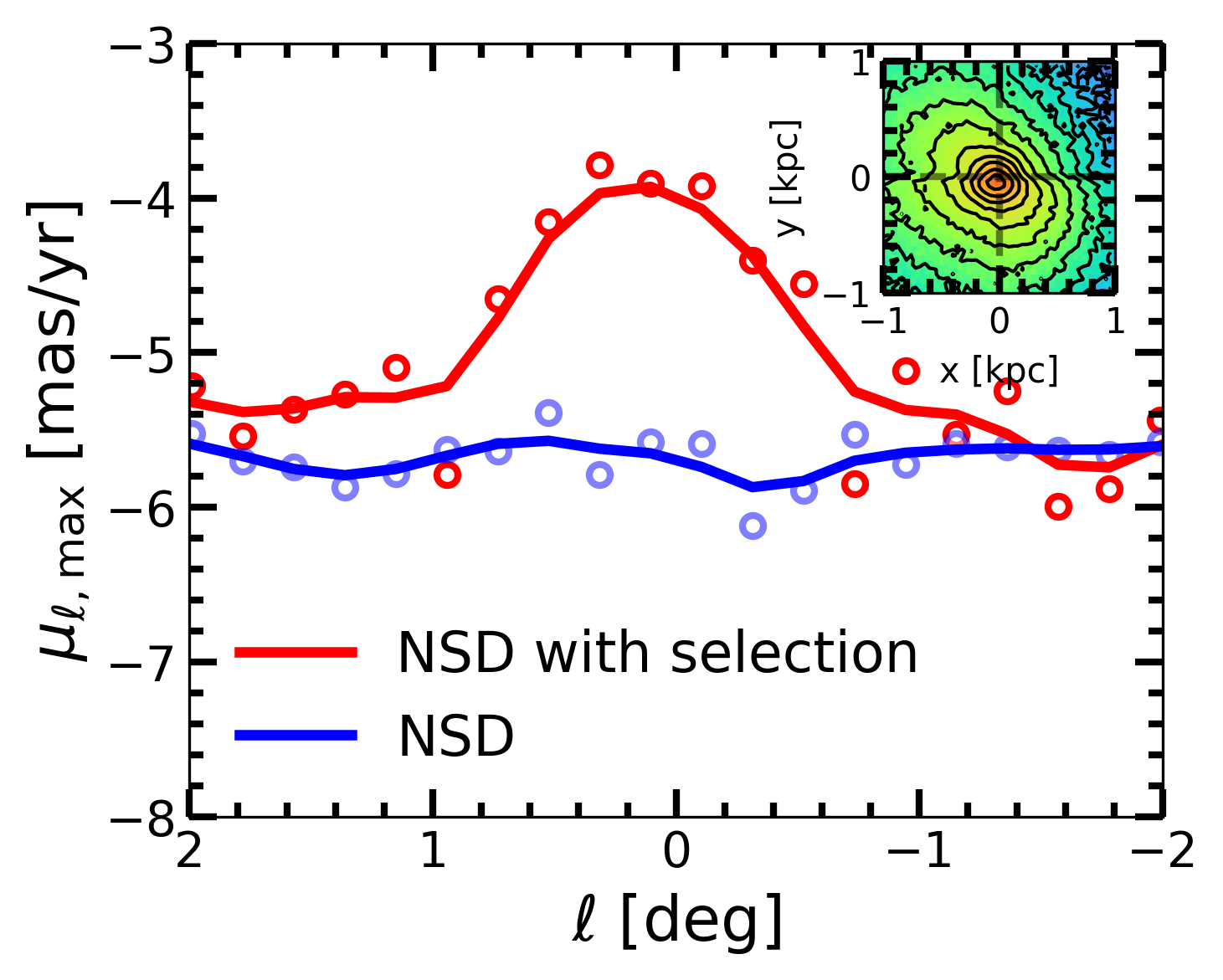}
    \caption{Distribution of $\mu_{\ell,\text{max}}$ vs $\ell$ for the NSD model without selection effects (blue markers and curve), and with the observational selection function applied (red). The inset shows the resulting face-on density distribution after selection. The distribution is no longer constant, but produces a strong peak in the distribution which could be misinterpreted as evidence for a nuclear bar if observational biases are not carefully accounted for.}
    \label{fig:axisymmetric_ext}
\end{figure}

The black open circles in Fig.~\ref{fig:mul_vs_l} show the $\mu_{l,max}$ distribution for stars sampled from individual fields in the midplane in order to emulate observations. In each field, we sample $N=1500$ stars from the S-NSB model, with uncertainties estimated via a bootstrapping approach. For each bin in $\ell$, we perform resampling of the stars with replacement for $1000$ iterations and the uncertainty  determined via Eq.~\ref{eqn:vertex_deviation_error}. 

In cases where $\alpha$ is small (e.g., $\alpha \lesssim 20\degree$), the step feature in the distribution is observable. However, as $\alpha$ becomes larger, it is unlikely that this method will reveal the presence of a nuclear bar. In addition, we note that even with $\sim 1500$ sources per field the distribution can be very noisy. Nonetheless, the sample sizes available in surveys in the near future will allow this diagnostic to be employed.

In Fig.~\ref{fig:axisymmetric_ext}, we test the impact of extinction on the $\mu_{\rm l,max}$ distribution. The blue markers/curve again show the distribution for the axisymmetric NSD. We then apply the selection function described in Sec.~\ref{sec:lv_ext} (see also Fig.~\ref{fig:selection}) to the NSD and re-compute the $\mu_{\rm \ell,max}- \ell$ distribution (red markers/curve). The inset in Fig.~\ref{fig:axisymmetric_ext} shows the density distribution of the NSD when the selection is applied. Extinction has a strong impact on the $\mu_{\rm l,max}\mhyphen \ell$ distribution, creating a strong peak which is slightly shifted towards positive longitude. If not properly accounted for, extinction can easily produce a signal mimicking that of a nuclear bar; a major drawback of this diagnostic.

\section{Discussion and conclusion}\label{sec:discussion}

In this study, we have employed N-body simulations to systematically investigate kinematic diagnostics for detecting a nuclear bar in the Galactic centre. The primary objective was to develop kinematic diagnostics that are both theoretically robust and observationally practical given current and future data. The models were constructed to test the difference in kinematic signatures between an axisymmetric nuclear stellar disc and a nuclear bar. We also tested the role of extinction due to the presence of dust and gas in the Galactic centre using a toy model. We assessed three methods for detecting a nuclear bar: (1) Vertex deviation of the velocity ellipse (Sec.~\ref{sec:vertex_deviation}), (2) the asymmetry in the $\mu_{\rm \ell}-\ell$ distribution (see Sec.~\ref{sec:mu_l_assymetry}), and (3) the correlation/anti-correlation between the Gauss-Hermite moment $h_3$ and the line-of-sight velocity, $V_{\rm los}$ (see Appendix~\ref{sec:gh_moment}).

Our study indicates that the vertex deviation, \lv, provides the most robust method of detecting a nuclear bar in the Galactic centre (see Sec.~\ref{sec:model_vdev}). An NSD produces a velocity ellipse with its major axis aligned with the Galactic longitude velocity (i.e, \lv$\approx \pm 90\degree$), while a nuclear bar deviates from this, producing an \lv\ value which depends on its angle, $\alpha$. A nuclear bar with its near side at negative longitudes produces positive \lv, which contrasts sharply with the background negative \lv\ of the large-scale bar. Conversely, a nuclear bar with its near side towards positive longitudes produces a negative \lv. Importantly, contamination of roughly $20\mhyphen25\%$ from the main bar minimally affects \lv\ in the central region where the NSD density dominates, which spans roughly $|\ell|<0.9\degree$ and $|b|<0.25\degree$. However, in cases where $\alpha=\pm 90\degree$, the \lv\ signals of an NSD and NSB are practically identical. 

We also explored the strong possibility that the MW's NSD is not fully axisymmetric, but elliptical, with its major axis orthogonal to the main bar, as is observed in external galaxies \citep[e.g.,][]{Bittner+2020}. We find that in such cases, \lv\ is positive (see Sec.~\ref{sec:orthogonal}), and mimics the kinematic signature of a nuclear bar viewed nearly side-on with its semi-major axis at negative longitude. Therefore, should a positive \lv\ in the Galactic Centre be detected in future studies an NSD can not be easily ruled out.

The uncertainties associated with \lv\ are a function of sample size and nuclear bar strength. In the weak bar case (W-NSB), stellar velocities are more isotropic (smaller $\beta$) (see Fig.~\ref{fig:lv_a2}), which systematically increases the uncertainty. Therefore, the sample size required for a statistically significant measurement of \lv\ changes with nuclear bar strength. In the strong NSB case (S-NSB), we find that roughly $N=500$ is sufficient, whereas this increases to $N=1,000$ for a weaker NSB. In this context, one of the key advantages of using vertex deviation is the ability to combine stars from different observational fields, increasing the statistical significance of the \lv\ measurement. This suggests that the data available in the current KMOS survey of the NSD \citep{Fritz+2021} may allow for the detection of a nuclear bar via the vertex deviation diagnostic. This possibility will be addressed in a companion paper.

We find that \lv\ could be impacted by extinction, however the effect is not strong. We applied a three-component probabilistic selection function that accounts for distance-dependent detectability, midplane extinction and crowding, and asymmetric dust extinction as a function of Galactic longitude (Appendix~\ref{sec:selection}). We find that this shifts \lv\ towards positive values, possibly mimicking an NSB with its near side towards negative Galactic longitudes. 

Similar to \citet{Fernandez+2025} for a large-scale bar, we find that while \lv\ reveals non-axisymmetry, it should not be considered as a measure of the \emph{level} of non-axisymmetry. Indeed, for a given nuclear bar angle, \lv\ will take on roughly the same value, regardless of bar strength (see Fig.~\ref{fig:lv_a2}). Instead, the anisotropy, $\beta$, of the velocity ellipsoid may be employed in order to characterise the bar amplitude, as it increases with bar amplitude.

The observed asymmetry in the $\mu_{\rm \ell}-\ell$ plane provides another method of detecting a NSB. In Fig.~\ref{fig:mul_vs_l} we show the distribution of $\mu_{\rm l,max}$ versus $\ell$ for various orientations (panels) of the NSB (red markers/curves). The NSB will induce deviations from the otherwise flat distribution produced by an axisymmetric NSD (blue points). This detection method is most effective in cases where the NSB angle is small ($-20\degree<\alpha<20\degree$). This produces a "step" feature in the inner $|\ell|<1\degree$. For NSB angles $|\alpha| > 30\degree$, the asymmetry (and therefore the central slope) becomes weaker, and this method becomes unreliable. In such cases, using the vertex deviation is preferred. Two important caveats are that unlike the vertex deviation, this method is highly sensitive to the asymmetric extinction in the Galactic centre (see Fig.~\ref{fig:axisymmetric_ext}). Secondly, while the bootstrapping we applied reflects how $\mu_{\rm \ell}$ varies depending on sampling, individual sources in observational data will have their own $\mu_{\rm \ell}$ uncertainties, further diluting the signal.

We also tested the correlation or anti-correlation between the Gauss--Hermite moment $h_3$ and the line-of-sight velocity $V_{\mathrm{los}}$. Although this method has proven effective for identifying NSDs in external galaxies \citep[e.g,][]{Lorenzo+2013}, we find no observable distinction between the NSD and NSB in our models: both exhibit similar $h_3$--$V_{\mathrm{los}}$ anti-correlation for all NSB orientations (see Appendix~\ref{sec:gh_moment}). Consequently, this diagnostic does not provide meaningful constraints on the presence of an NSB in the MW, and we therefore include the full analysis only in the Appendix for completeness.

Overall, our findings demonstrate that $l_v$ is both robust against extinction and selection effects and sensitive across a wide range of nuclear-bar orientations. The $\mu_\ell$--$\ell$ asymmetry method plays a complementary but limited role, while Gauss--Hermite diagnostics do not provide discriminating power in this context.

\section{Future work}\label{sec:future_work}

The number of stars required to reliably measure the vertex deviation is available in the KMOS survey \citep{Fritz+2021}, which contains $\approx 3000$ sources. This possibility shall be explored in a companion paper. Nevertheless, a detailed picture of the NSD region will likely require the larger sample sizes and precise measurements from future surveys such as the KMOS public survey, VVVX-GalCen (Nogueras-Lara et al. in prep). This will provide temperatures, velocities, spectral indices, and metallicities for $\sim 40\, 000$ sources in the NSD region. The MOONS REDdened Milky WAY (REDWAY) survey \citep{Cirasuolo+2020,Gonzalez+2020}, will also provide high resolution spectra for $\sim13,000$ stars within the region $|\ell|<1\degree$ and $|b|<0.5\degree$, enabling measurements of radial velocities and metallicities. Similarly, the JASMINE Galactic Centre survey (GCS) \citep{Kawata+2024} will cover a region $-0.7\degree<|\ell|<1.4\degree$ and $|b|<0.6\degree$, and is expected to provide accurate positions and proper motion measurements for about $1.2\times10^5$ stars. Lastly, the \textit{Roman} Galactic Bulge Time Domain Survey \citep{Terry+2023} will provide high-precision photometric and astrometric measurements of $\sim3.3$ million stars over a $0.28$ deg$^2$ field centred on SgrA*, with proper motion precisions of $2.5-3.5\, \mu$as/yr. These datasets will facilitate the application of the diagnostic tools discussed in this paper, advancing our understanding of the Galactic Centre.

\begin{acknowledgements}
    We thank the anonymous referee for providing valuable feedback which improved the manuscript. KF, MCS, ZF, XL, AV, and MD acknowledge financial support from the European Research Council under the ERC Starting Grant “GalFlow” (grant 101116226). MCS further acknowledges financial support from the Fondazione Cariplo under the grant ERC attrattivit\`{a} n. 2023-3014. JLS acknowledge support from the Royal Society (URF\textbackslash R1\textbackslash191555; URF\textbackslash R\textbackslash 241030).

    The simulations in this paper were run both at the High Performance Computing Facility of the University of Lancashire and at the DiRAC Shared Memory Processing system at the University of Cambridge, operated by the COSMOS Project at the Department of
    Applied Mathematics and Theoretical Physics on behalf of the STFC DiRAC High Performance Computing (HPC) Facility: www.dirac.ac.uk. This equipment was funded by the Department for Business, Innovation and Skills (BIS) National E-infrastructure capital grant ST/J005673/1, STFC capital grant ST/H008586/1, and STFC DiRAC Operations grant ST/K00333X/1. DiRAC is part of the National E-Infrastructure. FNL gratefully acknowledges financial support from grant PID2024-162148NA-I00, funded by MCIN/AEI/10.13039/501100011033 and the European Regional Development Fund (ERDF) “A way of making Europe”, from the Ramón y Cajal programme (RYC2023-044924-I) funded by MCIN/AEI/10.13039/501100011033 and FSE+, and from the Severo Ochoa grant CEX2021-001131-S, funded by MCIN/AEI/10.13039/501100011033.
    
\end{acknowledgements}

\bibliographystyle{aa}
\bibliography{refs} 

\begin{appendix}

\section{The metal-poor halo}\label{sec:metal_poor}
Observations reveal that while the Milky Way's boxy-peanut bulge is predominantly traced by metal-rich stars, the more spheroidal metal-poor ($\feh < -0.5$) population must include a contribution of $\sim10\%$ from the halo \citep[e.g.,][]{Ness+2013, Rojas-Arriagada+2020}. This metal-poor halo is not included in the large-scale bar simulation employed in this work and could therefore represent an additional source of contamination in our analysis.

To assess the potential impact of this population, we construct a synthetic metal-poor component by sampling velocities from Gaussian distributions with dispersions $\sigma_{\mathrm r} = 99 \kms$ and  $\sigma_{\mathrm \ell} = 138 \kms$ (Table 1 in \citet{Babusiaux+2010}), which characterizes the kinematics of metal-poor stars in the bulge. When sampling stars from our models to compute the diagnostic metrics, we add a $10\%$ contribution from this synthetic metal-poor population. We find that the resulting impact on our diagnostics is consistently within the uncertainties determined through bootstrapping (see Sec.~\ref{sec:error_handling}). Given its negligible effect on our results, we exclude this component from our final modelling framework.

\section{Vertex deviation derivation}\label{sec:vertex_deviation_der}

\begin{figure}
    \centering
    \includegraphics[width=\linewidth]{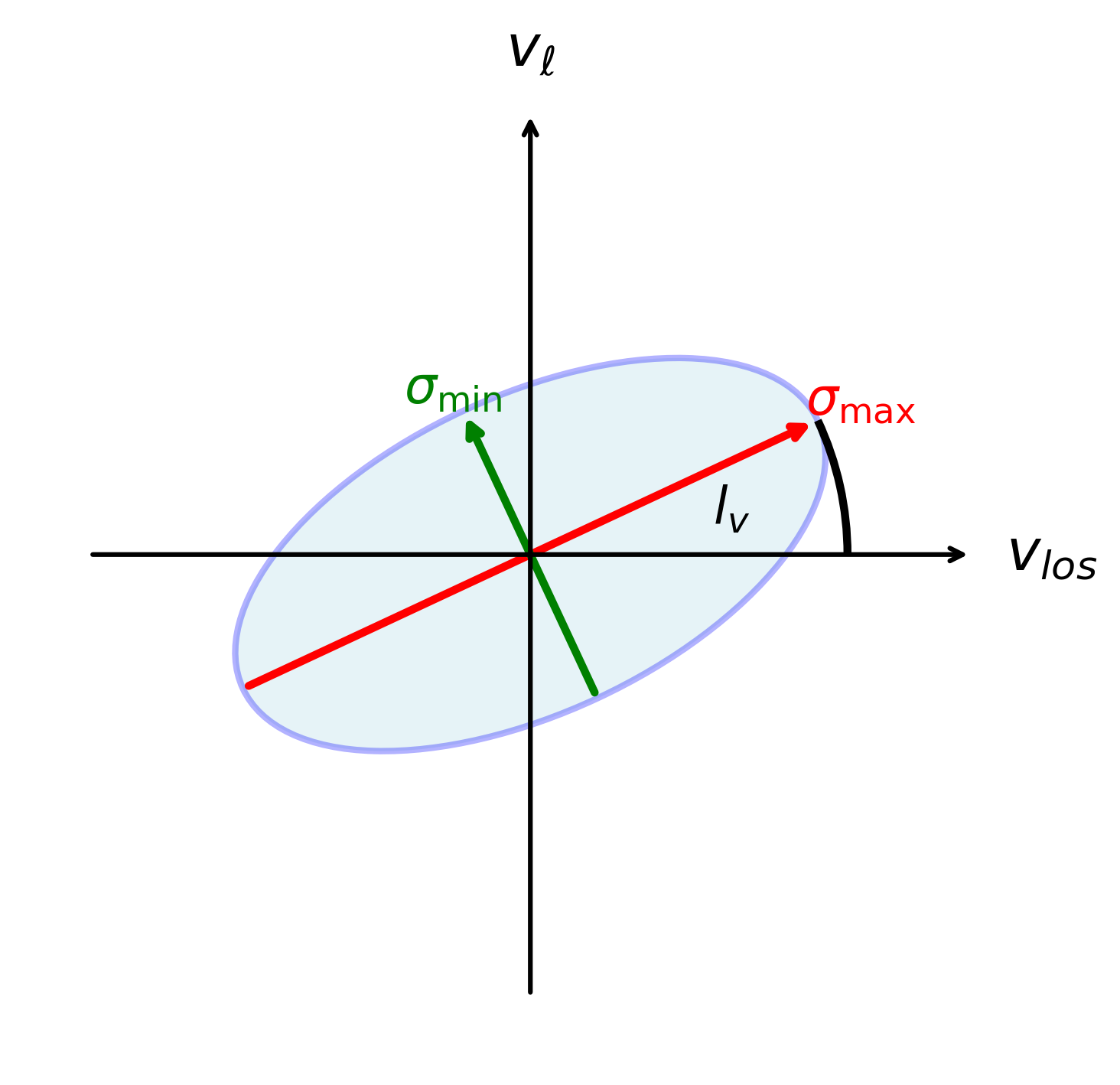}
    \caption{Schematic illustration of the velocity ellipse. The two eigenvectors (Eq.~\ref{eqn:eigenvectors}) of the velocity dispersion tensor (Eq.~\ref{eqn:cov_matrix}) are shown as green and red arrows. The vertex deviation (Eq.~\ref{eqn:vertex_deviation_ap}) is the angle indicated by the black arc.}
    \label{fig:schematic}
\end{figure}

In 2 dimensions, the velocity dispersion tensor defined by Eq.~\eqref{eqn:dispersion} can be written as:
\begin{equation}
    \renewcommand{\arraystretch}{1.5}
    \sigma_{ij} = 
    \begin{pmatrix}
        \sigma^2_{11} & \sigma^2_{12} \\
        \sigma^2_{21} & \sigma^2_{22}
    \end{pmatrix},
    \renewcommand{\arraystretch}{1}
    \label{eqn:cov_matrix}
\end{equation}
where $1$ refers to the first coordinate (in our case, the line-of-sight direction $v_{\rm los}$) and $2$ to the second coordinate (in our case, the transverse direction $v_\ell$), and the tensor is symmetric, \( \sigma^2_{12} = \sigma^2_{21} \).
Because it is symmetric, the dispersion tensor is always diagonalisable. The eigenvalues are given by:
\begin{equation}
    \sigma_{\mathrm{max/min}} = \frac{1}{2}\left[\sigma^2_{11}+\sigma^2_{22}\pm D\right],
    \label{eqn:eigenvalues}
\end{equation}
where the $+$ ($-$) sign corresponds to $\sigma_{\mathrm{max}}$ ($\sigma_{\mathrm{min}}$), and we have defined
\begin{equation}
    D \equiv \sqrt{(\sigma^2_{11}-\sigma^2_{22})^2+4(\sigma^2_{12})^2}.
    \label{eqn:D_def}
\end{equation}
The corresponding eigenvectors are:
\begin{equation}
  \mathbf{V}_{\mathrm{max/min}}  \propto \begin{pmatrix} 
    2\sigma^2_{12} \\
    \sigma^2_{22}-\sigma^2_{11}\pm D
    \end{pmatrix}.
    \renewcommand{\arraystretch}{1}
    \label{eqn:eigenvectors}
\end{equation}
The vertex deviation, denoted as \( l_v \) and illustrated by the black arc in Fig.~\ref{fig:schematic}, is defined as the angle that the eigenvector corresponding to the largest eigenvalue ($\mathbf{V}_{\rm max})$ makes with the $v_{\rm los}$ axis. Elementary trigonometry yields:
\begin{equation}
    \tan(2l_{\rm v}) = \frac{2\sigma^2_{12}}{\sigma^2_{11}-\sigma^2_{22}},
    \label{eqn:tan_2lv}
\end{equation}
and therefore:
\begin{equation}
\boxed{    l_{\rm v}= \frac{1}{2}\arctan\left(\frac{2\sigma^{2}_{12}}{\sigma^{2}_{11}-\sigma^{2}_{22}}\right)}
    \label{eqn:vertex_deviation_ap}
\end{equation}
The vertex deviation \lv\ takes values within the range \( [-90^\circ, 90^\circ] \) and represents the angle that the semi-major axis of the velocity ellipse forms with the $v_{\rm los}$ axis as illustrated in Fig.~\ref{fig:schematic}.

Note that sometimes a modified formulation is employed in the literature \citep[e.g.,][]{Soto+2007,Babusiaux+2010,Debattista+2019,Simion+2021,Fernandez+2025}, which incorporates the absolute value of the denominator:
\begin{equation}
    l_{\rm v}= \frac{1}{2}\arctan\left(\frac{2\sigma^{2}_{12}}{|\sigma^{2}_{11}-\sigma^{2}_{22}|}\right).
    \label{eqn:vertex_deviation_abs}
\end{equation}
In this case, \lv\ takes values in the interval \( [-45^\circ, 45^\circ] \) and is the angle that the semi-major axis of the velocity ellipse makes with the nearest coordinate axis (either $v_{\rm los}$ or $v_\ell$). We do not use this alternative definition in this paper.

\section{The SPH model}\label{sec:sph_model}

We employ an N-body+SPH model of an isolated disc galaxy to explore how an elliptical, orthogonal NSD impacts our results. This model is equivalent to model M3\_nc\_b in \citet{Fiteni+2021}, and has also been described extensively in \citet{Cole+2014} and \citet{Debattista+2017}. Therefore, we give here only a brief overview of the model. Initially, the model comprises of a corona of hot gas embedded within a dark matter halo with a virial radius of $r_{200} \simeq 200 \kpc$ and a virial mass of $M_{200} = 10^{12} \Msun$. Both the gas and dark matter are represented by $5 \times 10^6$ particles each, with softening parameters of  $\epsilon = 50 \pc$ and $\epsilon = 100 \pc$ for the gas and dark matter respectively. All stars form from the gas, with no stars present at $t=0$. The initial radial profile of the gas corona is identical to that of the dark matter halo, but contains only $10\%$ of its mass. Gas particles are assigned a tangential velocity that results in a spin parameter $\lambda \equiv J |E|^{1/2}/(GM_{\mathrm{vir}}^{5/2}) = 0.065$, where $J$ and $E$ denote the total angular momentum and the energy of the gas particles respectively, and $G$ is the gravitational constant. The conditions for star-formation to occur require that the cold gas ($T < 15,000$ K) density exceeds 0.1 cm$^{-3}$ and is part of a converging flow. Star particles forming from the cooling gas also have a softening parameter of $\epsilon = 50 \pc$. The star-formation efficiency is set to $5\%$, and we employ the feedback recipe described by \citet{Stinson+2006}. The supernova feedback model injects $10\%$ of the $10^{51}$ erg per supernova back into the interstellar medium. Additionally, we implement turbulent diffusion for gas mixing as outlined by \citet{Shen+2010}. The model was evolved for $10 \Gyr$ with {\sc gasoline} \citep{Wadsley+2004, Wadsley+2017}, and develops a long-lived primary bar early in its evolution at roughly $3\Gyr$. The orthogonal NSD structure forms later and is comprised mostly of young stars with ages $\tau_{\star} < 4 \Gyr$.

\section{Observational Selection Effects}\label{sec:selection}

To generate a realistic mock observation from the models, we applied a three-component probabilistic selection function that accounts for distance-dependent detectability, midplane extinction and crowding, and asymmetric dust extinction as a function of Galactic longitude. Each star in the model was assigned a detection probability based on these effects, and the final "observed" sample was generated through Monte Carlo sampling. The three components of our selection function are illustrated in Fig.~\ref{fig:selection}.

Due to extinction effects and crowding in the midplane, stars on the far side of the disc $(d > 8.2 \kpc)$ are less likely to be detected. We model this selection bias using a skewed normal distribution (Fig.~\ref{fig:selection}a):

\begin{equation}
P(d) = \frac{1}{N} \cdot \phi_{\rm skew}(d; \mu = 8.2\,{\rm kpc}, \sigma = 1\,{\rm kpc}, \alpha = -1)
\end{equation}

\noindent where $\phi_{\rm skew}$ is the skewed normal probability density function with location parameter $\mu = 8.2$~kpc, scale parameter $\sigma = 1$~kpc, and skewness parameter $\alpha = -1$. The negative skewness produces an asymmetric distribution that favours stars on the near side of the NSD ($d < 8.2$~kpc), mimicking the observational reality that nearer stars are more readily detected than those on the far side of the Galactic Centre. The distribution is normalized such that $\max(p_{\rm dist}) = 1$.

Severe stellar crowding and extinction significantly reduce detection completeness near $b = 0\degree$. We also model the completeness with latitude (Fig.~\ref{fig:selection}b):

\begin{equation}
P(b) = 1 - 0.7 \exp\left(-\frac{|b|}{b_{\rm scale}}\right)
\end{equation}

\noindent where $b_{\rm scale} = 0.1\degree$ is the characteristic latitude scale. This functional form imposes a maximum $70\%$ incompleteness at the Galactic plane ($b = 0\degree$), with completeness recovering exponentially toward higher latitudes. As visible in Fig.~\ref{fig:selection}b, the detection probability drops to only $30\%$ at the midplane and recovers to nearly full completeness by $|b| \sim 0.5\degree$.

Observations of the inner Galaxy reveal a strong asymmetry in dust distribution, with approximately two-thirds of the obscuring material concentrated at positive Galactic longitudes. We model extinction using a hyperbolic tangent profile (Fig.~\ref{fig:selection}c):

\begin{equation}
P(\ell) = 0.5 + 0.25 \left[1 - \tanh\left(\frac{\xi \cdot \ell}{\ell_{\rm scale}}\right)\right]
\end{equation}

\noindent where the extinction strength coefficient is:

\begin{equation}
\xi = \begin{cases}
3.0 & \text{for } \ell > 0° \\
1.0 & \text{for } \ell \leq 0°
\end{cases}
\end{equation}

\noindent and $\ell_{\rm scale} = 0.5°$ controls the transition sharpness. This formulation ensures that detection probability decreases monotonically as longitude increases.

The total detection probability for each star is computed as the product of the three independent components shown in Fig.~\ref{fig:selection}:

\begin{equation}
P_{\rm total}(d, \ell, b) = P(d) \times P(b) \times P(\ell)
\end{equation}

To achieve a target overall completeness $C_{\rm target}$, we normalize the probability distribution:

\begin{equation}
p_{\rm norm}(d, \ell, b) = \frac{p_{\rm total}(d, \ell, b)}{\langle p_{\rm total} \rangle} \times C_{\rm target}
\end{equation}

\noindent where $\langle p_{\rm total} \rangle$ is the mean probability across all model stars. The final observed sample is generated through Bernoulli trials: each star is retained if a uniform random variate $u \sim \mathcal{U}(0,1)$ satisfies $u < p_{\rm norm}$. For our fiducial model, we adopted $C_{\rm target} = 0.15\%$, yielding a sparse sample that reflects the extreme observational challenges of detecting individual stars in the crowded, heavily extincted Galactic Centre region. The actual achieved completeness may differ slightly from the target due to the stochastic sampling process and probability clipping.

\section{Gauss-Hermite moments}\label{sec:gh_moment}

Line-of-sight velocity distributions have been widely used as kinematic diagnostics of galaxy features \citep[e.g.,][]{Gerhard+1993, marel+1993, Chung+2004, Bureau+2005, Debattista+2005, Ianuzzi+2015, Li+2018, Zakharova+2024, Fraser+2025A&A}. Specifically, correlations and anti-correlations between $h_3$ ("skewness") and the mean line-of-sight velocity in edge-on discs provide insight into the underlying stellar orbital structure of different components. A positive $h_{3} - V_{\rm los}$ correlation typically indicates elongated orbits (indicative of a bar), whereas an anti-correlation suggests the presence of more circular orbits characteristic of a stellar disc \citep{Bureau+2005, Ianuzzi+2015, Li+2018, Zakharova+2024}. An anti-correlation is also associated with the presence of an NSD \citep{Chung+2004, Zakharova+2024, Fraser+2025A&A}.

\begin{figure}
    \centering
    \includegraphics[width=\linewidth]{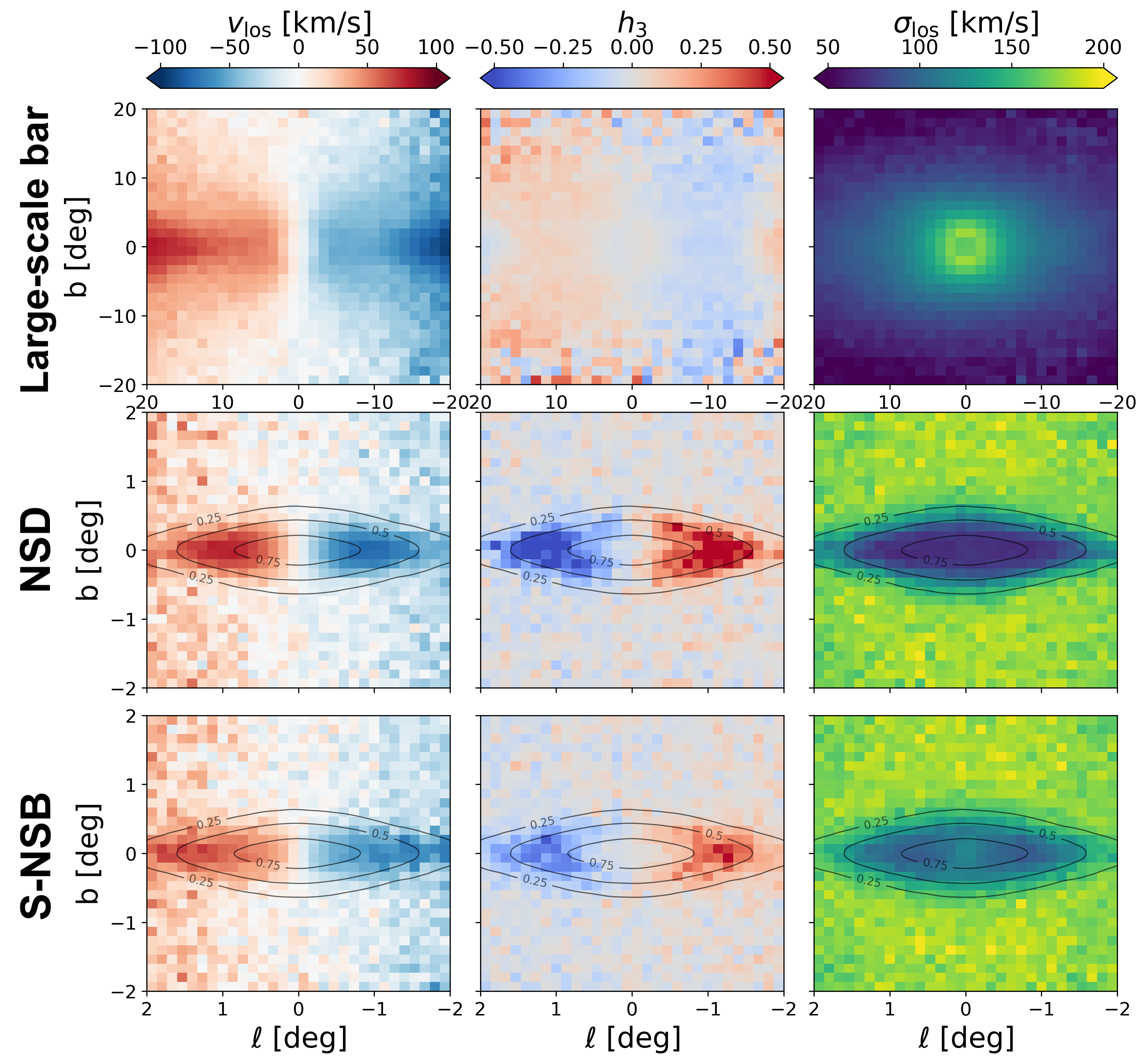}
    \caption{\emph{Left to right}: 2D maps of the line-of-sight velocity, the $h_3$ Gauss-Hermite moment, and the line-of-sight velocity dispersion. We show the distributions for the large-scale MW (top, note different scale), NSD (middle), S-NSB ($\alpha=90\degree$) models.}
    \label{fig:moments}
\end{figure}

In Fig.~\ref{fig:moments}, we plot 2D maps of the line-of-sight velocity (left), the $h_3$ Gauss-Hermite moment (middle), and the velocity dispersion (right). We show the distributions for the large-scale MW (top), NSD (middle), and S-NSB ($\alpha=90\degree$) models. In the S-NSB case, we tested the diagnostic for several values of $\alpha$, and found no variation among the results. We only show the results for $\alpha=90\degree$ as a representative case. As expected, we see positively correlated $h_3$ and $V_{\rm los}$ for the large-scale bar model (top row). Consistent with observations \citep[e.g.,][]{Chung+2004, Fraser+2025A&A}, the presence of the NSD (middle row) produces an anti-correlation between the mean velocity and $h_3$. In the bottom row, we show $V_{\rm los}$ and $h_3$ for the S-NSB model. We again find that these parameters are anti-correlated, indistinguishable from the NSD model case. This is consistent with \citet{Lorenzo+2013}, who carried out a detailed kinematical and stellar population analysis in a sample of four (SB0 - SBb) double-barred galaxies using integral-field spectroscopy. Their Figure 4 shows the face-on 2D maps of $h_3$ reveal that the distribution is not dependent on the nuclear bar angle, and is thus most likely related to the NSD in which the nuclear bar is embedded. The velocity dispersion (right column in Fig.~\ref{fig:moments}) also shows no clear observable difference between the NSD and S-NSB cases, save for a greater central dispersion for the S-NSB. Therefore, this method is unlikely to reveal the presence of nuclear bar in the Galactic centre.

\section{Vertical motion kinematics}\label{sec:vertical_motion}

\begin{figure}
    \centering
    \includegraphics[width=.95\linewidth]{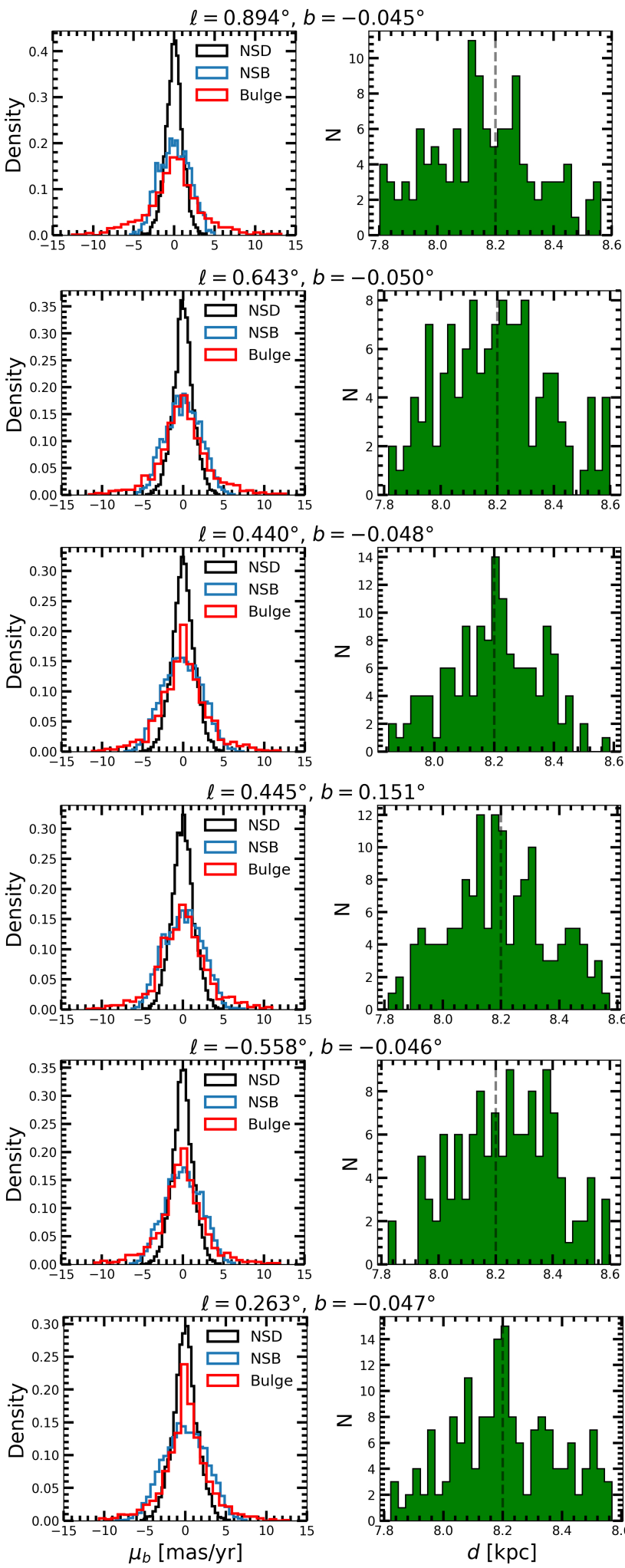}
    \caption{\emph{Left}: The vertical proper motion, $\mu_b$ for several fields. We show the distribution for the isolated NSD (black), isolated S-NSB (blue) and large-scale MW (red) models for stars in the distance range $7.8<d\, \mathrm{[kpc]} <8.4$. \emph{Right}: The distance distribution for stars excluded by the vertical cut $|\mu_b|<5\, {\rm mas/yr}$.}
    \label{fig:vertical_motions}
\end{figure}

The left column of Fig.~\ref{fig:vertical_motions} shows the $\mu_b$ distributions for stars within $7.8 < d\, [{\rm kpc}] < 8.4$ across various fields for the NSD, S-NSB, and large-scale bar models. The kinematically cooler NSD is characterised by low velocity dispersion, with typical values $|\mu_b| \lesssim 5\, {\rm mas\, yr^{-1}}$. The NSB shows moderately higher dispersion, reaching $|\mu_b| \approx 8\, {\rm mas\, yr^{-1}}$, while a subset of large-scale bar stars extends to significantly higher $|\mu_b|$ values. Applying a conservative cut of $|\mu_b| > 5\, {\rm mas\, yr^{-1}}$ across all fields reduces contamination from large-scale bar interlopers by at most $\sim2\%$, having negligible impact on our kinematic diagnostics. Nonetheless, observers may wish to apply such cuts when constructing clean samples of NSD stars. The right column of Fig.~\ref{fig:vertical_motions} shows the distance distribution of stars excluded by the  $|\mu_b| \lesssim 5\, {\rm mas\, yr^{-1}}$ cut. 

\end{appendix}

%
%

\end{document}